 \documentclass[a4paper]{article}

\usepackage[english]{babel}
\usepackage[utf8x]{inputenc}
\usepackage[T1]{fontenc}

\usepackage[a4paper,top=3cm,bottom=2cm,left=3cm,right=3cm,marginparwidth=1.75cm]{geometry}

\usepackage[dvipsnames]{xcolor}
\usepackage{amsmath}
\usepackage{graphicx}
\usepackage[colorinlistoftodos]{todonotes}
\usepackage[colorlinks=true, allcolors=blue]{hyperref}
\usepackage{authblk}
\usepackage{multicol}
\usepackage{bm}
\usepackage{color}
\usepackage{subfig}
\usepackage{hyperref}
\usepackage[normalem]{ulem}
\usepackage[square,numbers]{natbib}
\begin{document}
\title{An efficient implicit method for discrete dislocation dynamics simulations}

\author[1]{G\'abor P\'eterffy}
\author[1]{P\'eter Dus\'an Isp\'anovity}

\affil[1]{Department of Materials Physics, E\"otv\"os Lor\'and University}
\maketitle

\begin{abstract}
Plastic deformation of most crystalline materials is due to the motion of lattice dislocations. Therefore, the simulation of the interaction and dynamics of these defects has become state-of-the-art method to study work hardening, size effects, creep and many other mechanical properties of metallic specimens. Lot of efforts have been made to make the simulations realistic by including specific dislocation mechanisms and the effect of free surfaces. However, less attention has been devoted to the numerical scheme that is used to solve the equations of motion.
%Crystalline materials undergo irreversible plastic deformation as dislocations move in the crystal. The motion of the dislocations describes work hardening, size effects, creep and many other properties from the practical point of view, therefore modelling the motion of the dislocations is an important field of material sciences.

In this paper we propose a scheme that speeds up simulations by several orders of magnitude. The scheme is implicit because this type is the most efficient one for solving stiff equations that arise due to the long-range nature of dislocation interactions. The numerical results show that the method is not only faster than other approaches at the same numerical precision, but it can also be efficiently applied even without dislocation annihilation. The suggested method significantly increases the achievable volume and/or duration of discrete dislocation dynamics simulations and can be generalized for 3D simulations as well.

%The dynamics behind the motion of dislocations is complex because of the long range interactions between them through their anisotropic shear stress field. The result of the interactions is a stiff differential equation system which means explicit numerical methods can not solve it efficiently on big timescales. Regardless, the most often used numerical schemes to solve the equation system are from this category in 2 and 3 dimensions as well. This problem can be avoided by certain implicit methods, but on the contrary, with increasing number of dislocations it can become slower than the explicit method. The numerical results show, that the method achieved higher precision and magnitudes less runtime compared to other approaches. The method is not constrained by dislocation dipoles hence no dislocation annihilation is necessary. 

%The modified method can be used in 3D simulations as well, therefore its application can result in better simulations with larger simulated volume. 
\end{abstract}

\section{Introduction}
%The irreversible plastic deformation is an important topic of material sciences hence it is a priority to understand it better and deeper. These results are important during miniaturization processes mostly manufacturing electric parts but there are many other applications.

Although plastic deformation of crystalline materials is seemingly a smooth process on macroscopic scales, on the microscopic level it is characterized by intermittent local strain bursts \citep{Miguel2001, dimiduk_scale-free_2006}. The reason for this behaviour is that plastic deformation is the result of the motion of individual dislocations. Due to their long-range stress fields and complex short-range interactions these line-like defects can entangle forming a rigid network of dislocation lines. Upon external stress dislocations may locally disentangle and become mobile resulting in the accumulation of plastic strain. This and many other processes involving collective dislocation motion can be efficiently modelled by discrete dislocation dynamics (DDD) simulations \cite{bulatov_computer_2006}. They were successfully applied to various problems of materials science, such as describing the role of multi-junctions \cite{bulatov_dislocation_2006}, delivering a general picture of size effects \cite{el-awady_unravelling_2015} or to understand the statistics of strain burst sizes \cite{Csikor251}, just to mention a few.

%During plastic deformation of the crystalline materials macroscopic properties can be described by continuous functions, but they leap in an intermittent fashion on the microscopic scale \citep{Miguel2001}. This behaviour is emerging from the fact, that plastic deformation is a result of dislocation movement inside the crystal. These dislocations interact with each other through their long range stress field which leads to complex dynamics what can be simulated with the help of discrete dislocation dynamics (DDD) simulations. These simulations have an important role in the research of this field.

Several different kind of simulators are in use. The most important difference among them is the dimension of the simulation cell. Three dimensional (3D) approaches aim at modelling the full 3D dislocation microstructure and its dynamics. Due to the topological difficulties these simulations are numerically rather complex. Examples of such 3D DDD simulations include \mbox{ParaDiS} \citep{bulatov_dislocation_2006}, algorithm of Weygand \citep{WEYGAND2001420}, \mbox{microMegas} \citep{Devincre1745}, and Parametric Dislocation Dynamics \citep{ELAWADY20082019}. These simulations are used when the role of specific 3D mechanisms are investigated \citep{hussein_microstructurally_2015, cui_size-tuned_2018, lehtinen_multiscale_2016} or when one aims at a quantitative comparison with experimental results \citep{chang_multiscale_2010, greer_comparing_2008}. A much simpler representation of the dislocation systems is delivered by two dimensional (2D) simulations which only consider parallel straight dislocations. These are primarily used in cases when the physical consequences of long range dislocation interactions are investigated and when large simulation volumes, large ensembles for statistical averaging, longer timescales and/or higher numerical precision is required. Due to the simplification mentioned above a quantitative agreement with experiments cannot be expected, yet, these tools have been successfully applied to investigate, e.g., creep \textcolor{red}{\citep{miguel_dislocation_2002, ovaska_collective_2016}}, dislocation avalanches \citep{ispanovity_avalanches_2014, ovaska_quenched_2015, tsekenis_determination_2013} and patterning \citep{zhou_dynamic_2015, ispanovity_emergence_2017} of dislocations. An intermediate class is represented by 2.5D simulations, which are essentially 2D but with the inclusion of some 3D mechanisms, such as dislocation multiplication or dislocation pinning \cite{2004MSMSE..12..159B, Giessen1995, CLEVERINGA19973163, YEFIMOV2004279, kondori_discrete_2017, song_discrete_2019, lodh_microstructural_2019}.

%The second group has the dimension of "two and a half" which means these are actually simulations with the dimension of two, but 3D properties are also considered during the simulation (like dislocation sources, pinning sites, etc.) \citep{2004MSMSE..12..159B, Giessen1995} and the third, last group is where the simulations have the dimension of two.

%Based on the type of the examined problem different kind of simulators can be used. Several interesting results were published about micropillars and dislocation avalanches previously and the dimension of the simulators covered all categories. As an example, the 3D algorithm of Weygand was used in \citep{MOTZ20091744, SENGER20112937, Ispanovity2013, IMRICH2014240, PhysRevLett.105.085503, Csikor251} while considerable number of results were published from 2D \citep{Miguel2001, PhysRevLett.105.085503, PhysRevLett.112.235501, Ovaska2015} and 2.5D simulations as well \citep{papanikolaou_obstacles_2017}. ParaDiS was the first 3D simulator which simulated the formation of dislocation multi-junctions \citep{Bulatov2006}. Composite materials are often studied with simulations with dimension of "two and a half" \citep{CLEVERINGA19973163, YEFIMOV2004279}.  Pure 2D simulator's most common application is to study statistical properties \citep{Beato2005}, grain boundary effects \citep{PhysRevLett.114.015503}, aging \citep{PhysRevLett.98.075701} and relaxation processes \citep{PhysRevLett.107.085506}.

As it will be discussed in this paper the arising differential equation system that describes the dynamics of dislocations or dislocation segments is stiff. In such cases explicit methods are inefficient, because the achievable time-step is limited by the shortest timescale in the system, which is determined by the shortest dislocation dipole. This typically leads to a very slow propagation of the simulations even when the system is stationary. Nonetheless, most of the simulators mentioned above employ explicit methods \citep{Giessen1995,  papanikolaou_obstacles_2017, Sills_2014,devincre2011modeling}. This issue can be solved by implicit schemes \cite{Sills_2014}. Sills \emph{et al.}\ showed that with the help of the simple implicit trapezoidal method and the Newton-Raphson non-linear solver, in certain scenarios a speedup is achievable compared to the default Heun-method of ParaDiS if a sparse matrix was used. This was achieved by taking into account only short range elastic interactions in the Jacobian, defined with a fixed limit on the distance \cite{Sills_2014}. Gardner \emph{et al.} applied diagonally implicit Runge-Kutta methods and concluded that with high number of dislocation segments the gain in the stepsize compared to explicit methods was not large enough to compensate for the increase in the time needed to calculate one step \cite{gardner_implicit_2015}. Both of these studies were done using ParaDiS or DDLab (which is the serial version of ParaDiS written in MATLAB).

%and the implicit methods become inefficient for large system sizes because of the nonlinear equation system which has to be solved. The equation system's size scales with the number of the dislocations. This is one of the main problems of DDD simulations.

In this paper we present an efficient implicit integration scheme that can speed up simulations with several orders of magnitude while the numerical precision is the same or even higher than in case of explicit methods. The proposed scheme is capable of handling short dislocation dipoles efficiently that are responsible for the slowing down when using explicit methods. Another important property of the proposed scheme is that it can be tuned by a scale parameter which makes it possible to increase the efficiency of the simulator in different regimes of activity.

The method is tested on 2D edge dislocation systems because it is conceptually one of the simplest dislocation models. This simplicity makes it easier to compare the performance of different methods more accurately and precisely. However, the proposed method can be applied in more complicated systems or in simulations with higher dimension as well.

%This is not a new topic, publications and applications regarding implicit numerical schemes in DDD are present. \cite{Sills_2014} presented that with the help of the simple implicit trapezoidal method and the Newton-Raphson non-linear solver, in certain scenarios a speedup is achievable compared to the default Heun-method of ParaDiS if a sparse matrix was used. This was achieved by taking account of the closed range elastic interactions only for the Jacobian, with a hard, stationary limit on the distance. \cite{gardner_implicit_2015} applied diagonally implicit Runge-Kutta methods where they concluded that with high number of dislocation segments the gained larger stepsize needs more time to calculate than an explicit method would need to compensate for the stepsize loss. Both of these researches were done in ParaDiS or in DDLab (which is the serial version of ParaDiS in MATLAB.).

The structure of the paper is as follows: first, the 2D dislocation model and its background is introduced in section~\ref{sec:model_background}. Then, in section~\ref{sec:num_dif} the numerical difficulties arising from the properties of dislocation stress fields are discussed in detail followed by demonstrating the difference in performance between an explicit and an implicit method on the case of a single dislocation dipole. In section~\ref{sec:scheme_intro}, the new numerical method is presented and it is also explained how the previously highlighted features (better precision, larger stepsize, etc\dots) can be achieved. Then, a short insight is given into the actual implementation in section~\ref{sec:implementation} followed by the presentation of the numerical results on the efficiency in section~\ref{sec:numerical_results}. Section \ref{sec:summary} concludes the paper by presenting a summary of the results and our future plans.

%\textcolor{green}{
%\begin{itemize}
%\item mi az a diszlokáció dinamika? 
%\item milyen diszlokáció dinamikai szimulációk vannak, 2d, 2.5d, 3d hivatkozásokkal, mire jó?
%\item (probléma felvetés) numerikus problémák, explicit, implicit (pl. wei cai), melyik milyen %integrálást használ? implicit előnye/hátránya, időlépés, stb. (lehet 2 bekezdés is)
%\item miről szól a cikk, miért 2D-t csinálunk.
%\item fejezetenként mi lesz
%\end{itemize}
%}
\section{Discrete dislocation dynamics\label{sec:model_background}}

\subsection {Model description}
In the model considered, only straight and parallel edge dislocations are present with parallel slip planes. This system is effectively two-dimensional (2D) since it is enough to track the dislocation positions on a plane perpendicular to the dislocation lines. Let the $x$ axis of the 2D coordinate system be parallel with the Burgers vector $\bm b$ of the dislocations. The mechanical shear stress field of an individual dislocation in this case can be described with \citep{hirth_theory_1982}
\begin{align}
\tau_\text{ind}(x, y)&=\frac{\mu b}{2\pi(1-\nu)}\frac{x(x^2-y^2)}{(x^2+y^2)^2}\label{eq:field}
\end{align}
where $\mu$ and $\nu$ are material dependent elastic parameters whereas $x$ and $y$ are the distances from the dislocation along the corresponding axes.

For the motion of dislocations usually a linear velocity-force relationship is assumed. This corresponds to overdamped dynamics and is argued to be valid due to strong phonon drag acting on dislocations \citep{hirth_theory_1982}. The equations of motion in this case are
\begin{align}
\dot x_i (t)& = f_i(x_1(t), \dots, x_N(t), y_1, \dots, y_N) \nonumber \\
&= B^{-1} s_i b\left(\sum_{j=1, j\neq i}^Ns_j\tau_\text{ind}(x_i(t)-x_j(t), y_i-y_j) + \tau_\text{ext}\right) \label{eq:x_eq_motion}\\
\dot y_i(t) &= 0 \label{eq:y_eq_motion}
\end{align}
where $B$ is the dislocation drag coefficient and the indices $i$ and $j$ refer to the $N$ different dislocations. Symbol $s$ is equal to $1$ or $-1$ depending on the direction of the Burgers vector. The Burgers vector's length $b$ is the same for all dislocations. $\tau_\text{ext}$ represents external loading or any internal static stress field. For simplicity, in the current paper the focus is on the integration scheme, so, $\tau_\text{ext}=0$ is considered, but the generalization to $\tau_\text{ext}\neq0$ is rather straightforward. A square-shaped simulation area with periodic boundary conditions is employed in order to mimic an infinite medium. It is noted, that periodic dislocation images can be taken into account by using a modified stress field instead of $\tau_\text{ind}$ \citep{bako_dislocation_2006}.

%\textcolor{WildStrawberry}{\sout{This influences the stress-field as well, however, close to the origin $\tau_\text{ind}$ is not affected \textcolor{red}{[?]}} With an accordingly modified $\tau_\text{ind}$ this property can be taken into account \textcolor{red}{[?]}}.

% but it has to be pointed out that it can be applied in cases where other particles are present as well (e.g. solutes)

This model is one of the simplest representations of a complex dislocation network. Due to its assumptions, it is incapable for investigating many deformation phenomena related to, e.g., dislocation curvature or multiple slip. Its advantage is, on the other side, that it is conceptually much simpler than its 3D counterparts which makes numerical investigations faster and more reliable leading to the possibility to study larger systems and larger statistical ensembles at once with preserving high numerical precision. Another advantage is, opposed to 3D simulators where the same simulation can lead to different results based on loading and other factors \citep{Sills_2014, gardner_implicit_2015}, that the results in this case are reproducible and consistent. Consequently, several fundamental properties related to the long-range internal stresses of dislocations systems have already been successfully investigated with this model \citep{zaiser_statistical_2001, laurson_dynamical_2010, ispanovity_submicron_2010, tsekenis_dislocations_2011, salmenjoki_machine_2018}.

\subsection{Dimensionless variables}
\label{sec:dimensionless}

In the rest of this paper, in accordance with the $1/r$-type scale-free interaction $\tau_\text{ind}$, distance, stress and time are measured in units of $\rho^{-1/2}$, $\frac{\mu b \rho^{1/2}}{2\pi(1-\nu)}$ and $\frac{2\pi(1-\nu)B}{\mu b^2 \rho}$, respectively, where $\rho=N/L^2$ is the total dislocation density and $L$ is the size of the simulation area.

\subsection{Relaxation simulations}

\begin{figure}[ht]
    \centering
    \begin{picture}(0,0)
    \put(10,125){\sffamily{(a) Unrelaxed}}
    \put(145,125){\sffamily{(b) Relaxed}}
    \put(285,125){\sffamily{(c) Zoomed}}
    \end{picture}
    \includegraphics[width=0.95\linewidth, trim={0 1.2cm 0 1cm}, clip]{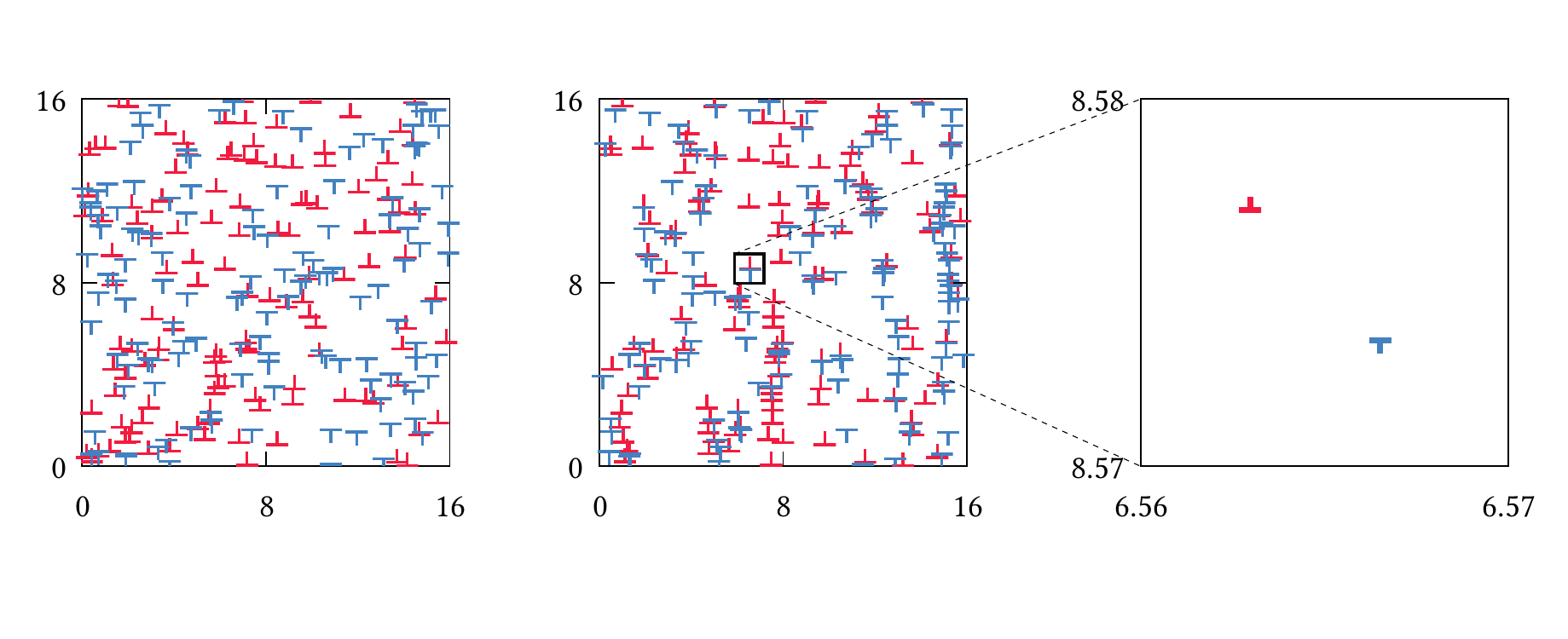}\\\vspace{1cm}
    \begin{picture}(0,0)
    \put(30,180){\sffamily{(d) Relaxation}}
    \end{picture}
    \includegraphics[width=0.7\linewidth]{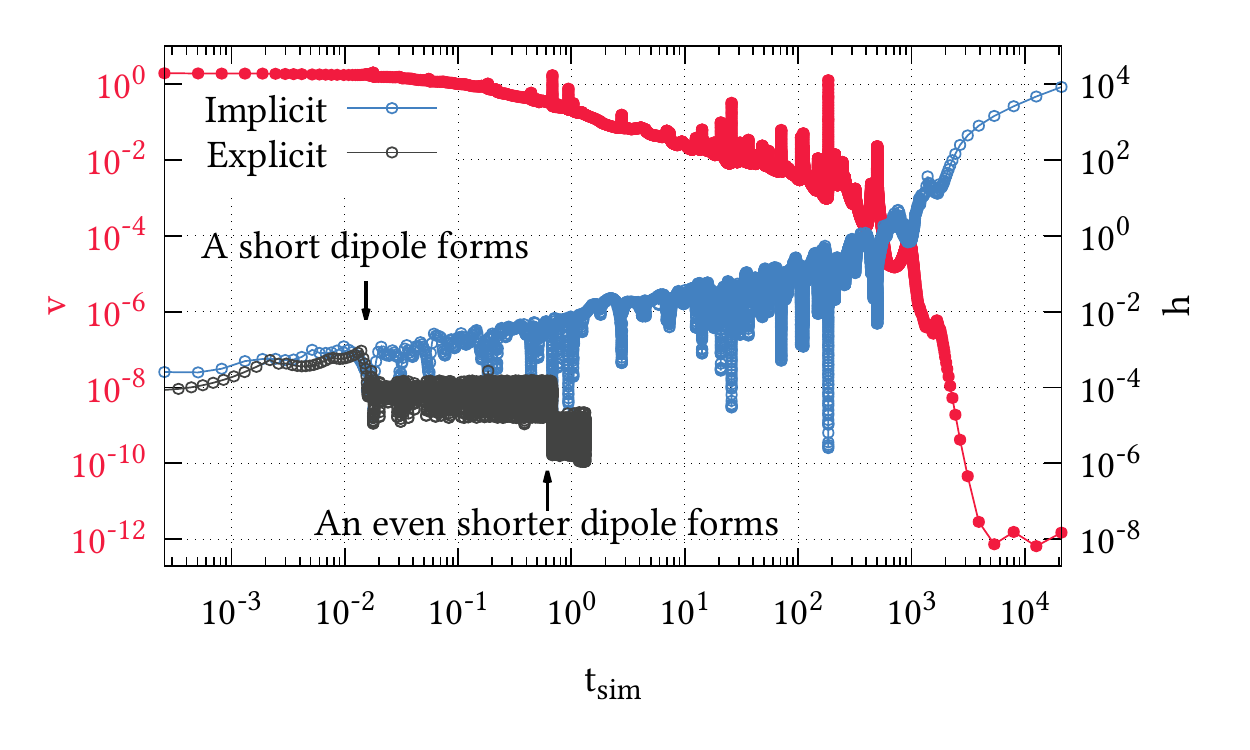}%relax_curve_256_id_00020_inf_exp.pdf}
    \caption{Relaxation of an initially random dislocation configuration with $N=256$ dislocations. (a): The initial random configuration. (b): The final relaxed configuration. (c): The relaxed configuration are characterized by spatial correlations. For instance, if opposite sign dislocations get close to each other they form a dipole as illustrated in the zoomed figure. (d): Average dislocation speed  $v$ (left scale, red colour) is shown as a function of simulation time $t_\text{sim}$. The relaxation is characterized by strong velocity fluctuations which eventually cease and the average velocity drops with 12 orders of magnitude. The numerical stepsizes $h$ (left scale) in case of an explicit 4.5th order Runge-Kutta (RK45) and the weighted implicit trapezoid scheme (WITS) method are also plotted. Whereas the timestep gradually increases during the relaxation for the implicit scheme, for the explicit case it is influenced only by the shortest dislocation dipole in the system. (It is noted, that the datapoints for the explicit method are restricted to small $t_\text{sim}$ values because of the significantly increased costs of computation.)}
    \label{fig:relaxation_figure}
\end{figure}
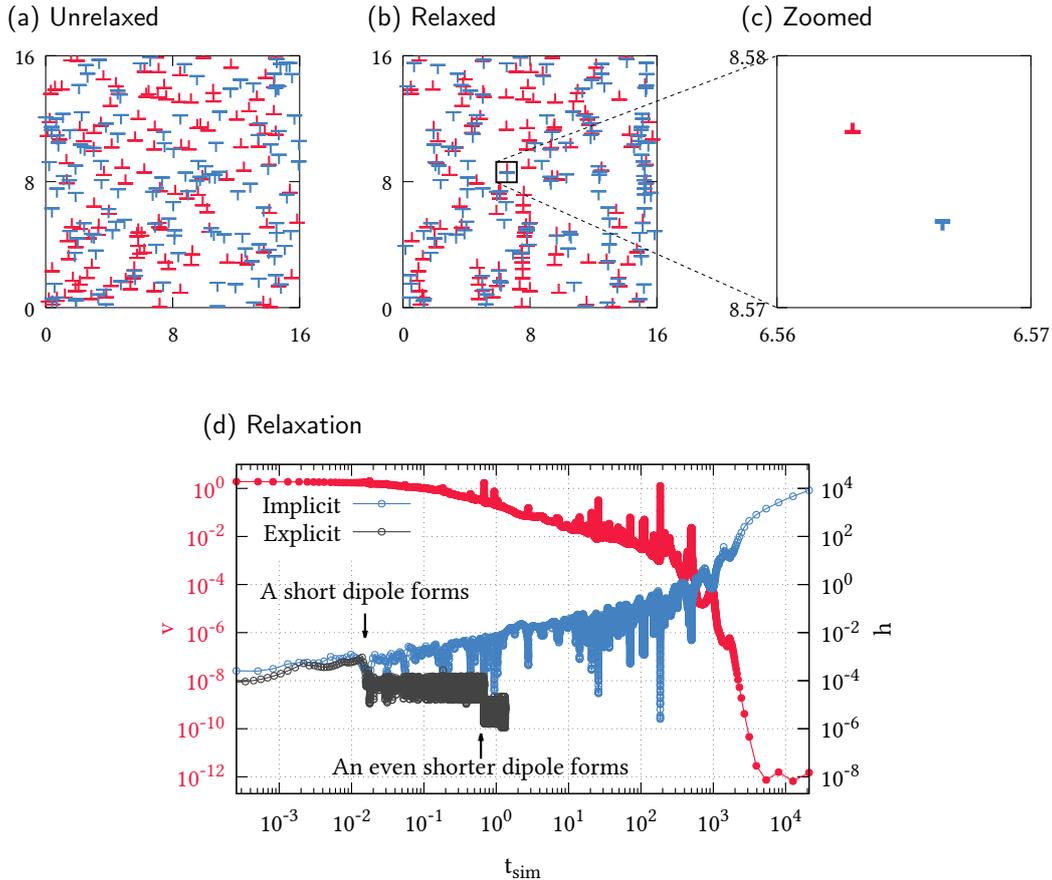

In the simulations we consider $N$ dislocations with zero net Burgers vector, i.e., $\sum_{i=1}^N s_i = 0$. The size of the simulation cell in the dimensionless units is then $\sqrt{N} \times \sqrt{N}$. Both $x$ and $y$ components of the initial coordinates of the dislocations are independent uniformly distributed random variables in the interval $[0, \sqrt{N}]$ [Fig.~\ref{fig:relaxation_figure}(a)]. Then the equations of motion (\ref{eq:x_eq_motion},\ref{eq:y_eq_motion}) are solved with zero external stress ($\tau_\text{ext}=0$). This leads to an initially rapid motion of dislocations which later, due to the strong dissipation introduced by the overdamped dynamics, slows down as the system gradually approaches an equilibrium state. During this process strong spatial correlations in the dislocation coordinates build up, corresponding to the build-up of low energy dislocation structures like dislocation dipoles and dislocation walls [Figs.~\ref{fig:relaxation_figure}(b) and \ref{fig:relaxation_figure}(c)] \citep{zaiser_statistical_2001}. The relaxation itself is a slow, scale-free process, and the relaxation time strongly increases with the number of dislocations considered \citep{ispanovity_criticality_2011}.

\section{Numerical difficulties\label{sec:num_dif}}

\subsection{Long-range interactions}

Due to the $1/r$-type long-range stress field $\tau_\text{ind}$ one cannot apply a cut-off in the mutual interactions. So, all the terms in the sum of Eq.~(\ref{eq:x_eq_motion}) have to be taken into account irrespective of the relative distance of the dislocations. Introduction of a cut-off is known to lead to the appearance of artificial dislocation patterns with a length scale proportional with the cut-off radius. \citep{gulluoglu_dislocation_1989}

The time complexity of every timestep is, therefore, $\mathcal O(N^2)$. In practice this means that if the linear extension of a simulation cell size is doubled (i.e.,~the number of dislocations is increased by a factor of 4) than computing a single timestep lasts 16 times longer. This extreme increase in computational cost makes investigation of large configurations rather difficult.

One possible solution to overcome this difficulty is the fast multipole method first introduced in 3D DDD simulations \citep{arsenlis_enabling_2007, zhao_new_2010, yin_computing_2012}, then also used successfully in the present 2D set-up \citep{bako_dislocation_2006, Isp_novity_2011}. This method reduces the time complexity to approximately $\mathcal O(N \ln(N))$. In this paper we do not address this issue any further and will use the exact summation of Eq.~(\ref{eq:x_eq_motion}) in the simulations.

\subsection{Stiff equations: limitations on time step}

The previous section was about how to evaluate the sum in Eq.~(\ref{eq:x_eq_motion}). Now we continue with the numerical scheme that is needed to perform the integration of Eq.~(\ref{eq:x_eq_motion}). 
The average velocity of the dislocations can be used to monitor the activity of a system during a simulation. In Fig.~\ref{fig:relaxation_figure}(d) the average velocity
\begin{align}
	v(t) &= \frac 1N \sum_{i=1}^N |v_i(t)|
\end{align}
is seen during the course of a representative simulation run with the full implicit weighted trapezoid method (abbreviated as WITS, described later in Sec.~\ref{sec:weight_selection}). As it is seen, dislocation activity strongly fluctuates but gradually ceases as the system approaches equilibrium. Although the average velocity has dropped in this case around 12 orders of magnitude which is close to the precision of the used double data type, it does never reach absolute zero due to some numerical noise always present in the system (the level of which can be controlled by the tolerance parameter of the adaptive stepsize routine down to the used architecture's limit).

One of the most common methods for dislocation simulations is the 4.5th order explicit Runge-Kutta scheme with adaptive stepsize control (abbreviated as RK45). This method is an explicit scheme (thus, relatively easy to implement) and its popularity in solving a system of ordinary differential equations is due to its high accuracy and relatively low computational cost of performing a single timestep.
%On the other hand, the previously mentioned implicit method is not a commonly used type because of the higher computational cost and complexity of the algorithm.

In Fig.~\ref{fig:relaxation_figure}(d) time stepsizes as a function of the simulation time for both mentioned methods are also shown. In the beginning of the simulation the actual time stepsizes for the RK45 method vary due to the adaptive stepsize control (being smaller in active periods and larger in quiescent regions) but with the formation of a short dipole it quite quickly settles at a quite low approximately constant value. At $t_\text{sim}\approx 0.6$ an even shorter dipole forms, so the timestep drops further and remains at the same value for the rest of the simulation. On the other hand, the timestep of the implicit method increases throughout of the simulation run with a total increase of more than 7 orders of magnitude. While it cannot be seen in the figure, the timestep is strongly affected by the tolerance parameter: better precision (lower tolerance) decreases the timestep and leads to longer simulation runs in real time.

The peculiar behavior of the timestep shown above is due to the \emph{stiffness} of the governing equations. This term applies to systems of ordinary differential equations where the timescales of the parallel processes are on a broad scale. The timestep of explicit methods follows the smallest timescale in the system, so for stiff systems the usage of implicit methods is advised. In the case of dislocations the stiffness is due to the $1/r$-type long-range interactions, since relaxation is very fast for nearby dislocations and slow down with increasing mutual distance. In the next section this will be quantified for a single dislocation dipole.

\subsubsection{Demonstration on a dislocation dipole \label{sec:dipole}}

\begin{figure}[ht]
\centering
\includegraphics[width=0.5\linewidth]{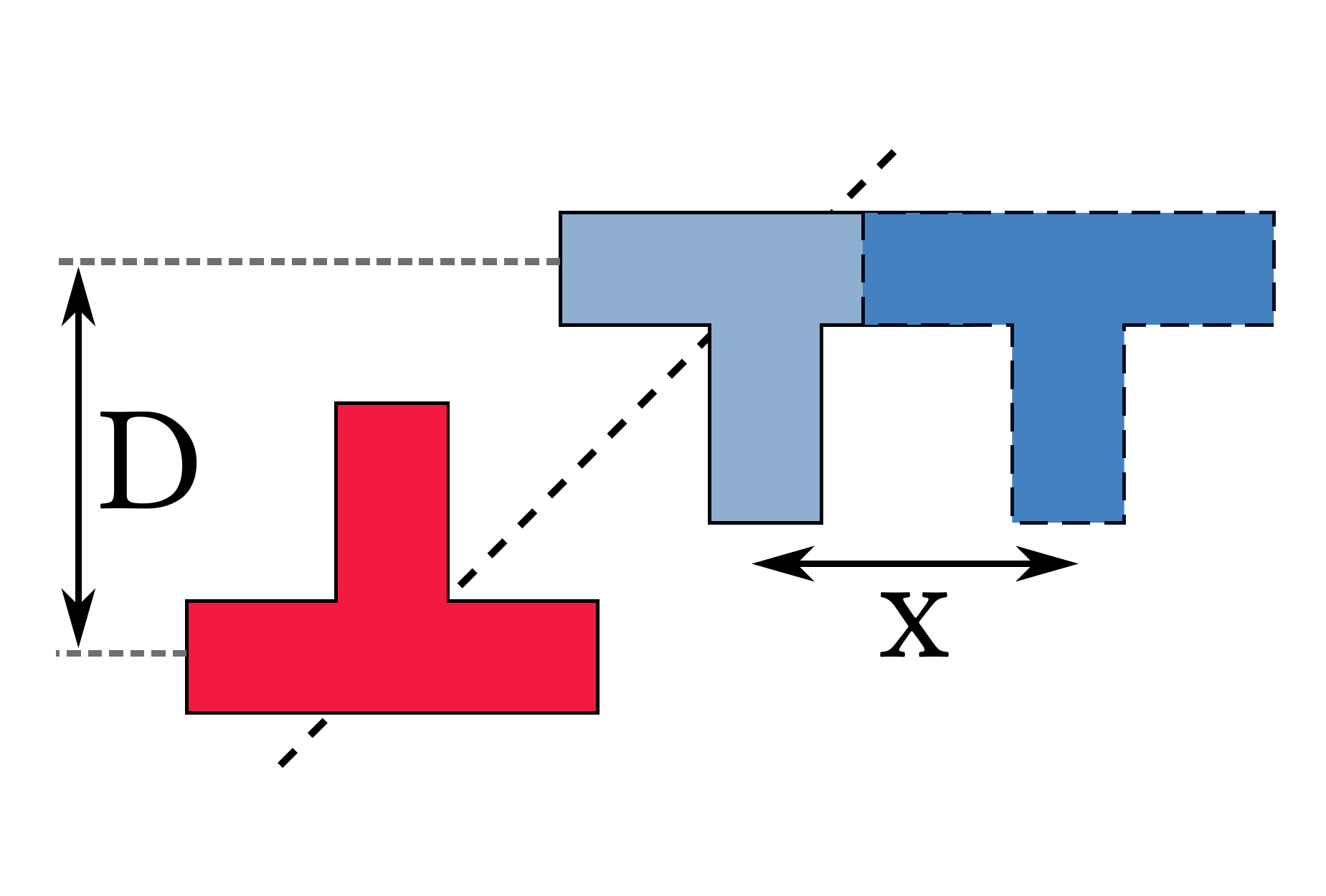}%dislocation_dipole_series_expansion.pdf}
\caption{A dislocation dipole that is used for the stability analysis of the different numerical methods. $D$ represents the size of the dipole while $x$ denotes the distance from the equilibrium position.\label{fig:dipole}}
\end{figure}

To understand the basic reason for the stepsize behaviour observed above and the difference between the explicit and the implicit numerical methods applied to the problem, a small dipole and its dynamics is considered. Such dipoles form in a large number during the simulations, and ones with small distances have the highest probability \citep{Csikor2004}. The considered setup can be seen in Fig.~\ref{fig:dipole}: the distance of the glide planes is $D$, and one of the dislocations is slightly shifted from the equilibrium $45^\circ$ position.

After the Taylor-expansion of the equation of motion Eq.~(\ref{eq:x_eq_motion}) for this dipole, in the $x\ll D$ limit one concludes for the top dislocation that
\begin{align}
\dot{x}=-\frac{1}{D^2}x. \label{eq:dipole_limit}
\end{align}
Here the dimensionless units of Sec.~\ref{sec:dimensionless} are used and it is noted that similar equation holds for the bottom dislocation as well. The solution of this equation is an exponential relaxation to $x=0$ with time constant $\tau = D^2$, that is, the relaxation timescale tends to zero quadratically for small dislocation dipoles.

%When two dislocations are present and they form a dipole what is shown by Fig.~\ref{fig:dipole}, $\dot x$ can be approximated by for one of them with Eq.~(\ref{eq:dipole_limit}) which can be obtained from the Taylor series based on Eq.~(\ref{eq:x_eq_motion}). For the other dislocation by definition the sign changes. The material dependent properties has been chosen as $2\pi$ hence the factors disappeared.

The different numerical integration schemes that can be used to solve Eq.~(\ref{eq:dipole_limit}) can be categorized as explicit or implicit methods. They can be sorted by inspecting the equation
%\begin{align}
%\frac{x(t+h)-x(t)}{h}&=F\left(x(t+h), x(t), t\right) \label{eq:derivative}
%\end{align}
\begin{align}
\frac{x^{k+1}-x^k}{h}&=F\left(x^{k+1}, x^k, x^{k-1}, \dots, t^k\right), \label{eq:derivative}
\end{align}
where $x^k$ approximates the solution $x(t^k)$ at $t^k=kh$, and $h$ is the timestep considered. If $F$ does not depend on $x^{k+1}$ then the method is an explicit method. Typical examples are the 4th order Runge-Kutta method or the simplest, forward Euler-method. If $F$ is dependent upon $x^{k+1}$ it is an implicit method and usually a system of non-linear equations has to be solved at every time step. In the following we investigate how the simplest explicit (Forward Euler) and implicit (Backward Euler) methods handle the above introduced perturbed dipole.

\paragraph{Forward Euler method}

The right-hand-side of Eq.~(\ref{eq:dipole_limit}) is here evaluated at timestep $k$, i.e., $F\left(x^{k+1}, x^k, x^{k-1}, \dots, t^k\right) = -x^k/D^2$. By rearranging Eq.~(\ref{eq:derivative}) one obtains
\begin{align}
%x^{k+1} &= x^k+h\dot{x}^k\label{eq:forward_euler}\\
x^{k+1} &= x^k\left(1-\frac{h}{D^2}\right). \label{eq:forward_euler_applied}
\end{align}
It is seen, that if the time step $h$ is larger than $h_\text{crit} = 2D^2 = 2\tau$ then instead of converging to $x=0$ the dislocation will diverge in an oscillatory way (that is, $|x^{k+1}| > |x^k|$ for $h > h_\text{crit}$). This happens irrespective of how close the dislocation is to the equilibrium position.

%By using the explicit forward Euler formula Eq.~(\ref{eq:forward_euler}), we get Eq.~(\ref{eq:forward_euler_applied}).  

\paragraph{Backward Euler method}

In contrast, $F\left(x^{k+1}, x^k, x^{k-1}, \dots, t^k\right) = -x^{k+1}/D^2$ is assumed when the backward Euler formula is used, and one gets
\begin{align}
x^{k+1}&=\frac{x_i^k}{1+\frac{h}{D^2}}. \label{eq:backward_euler_applied}
\end{align}
This means there is no limitation for $h$ in terms of convergence, dislocations will approach the equilibrium position $x=0$ independently large $h$ is (that is, $|x^{k+1}| < |x^k|$ for every $h$). This means this scheme is stable for the given problem.

\paragraph{Comparison of the two methods}
The forward Euler method is thus seriously limited in stepsize $h$ when a dipole forms, no matter what else happens in the system. The maximum allowed stepsize scales with the $D_\text{min}^2$, where $D_\text{min}$ is the characteristic distance of the shortest dipole in the system. This is also true, when this particular dipole is already in equilibrium. This can be observed in Fig.~\ref{fig:relaxation_figure} for a more complicated explicit scheme. Hence, the common solution in case of explicit methods is that under a certain threshold these dipoles are annihilated. Although this process also happens in real crystals, in simulations in favor of simulation speed often unrealistically large annihilation distances are applied that may lead to numerical side effects. In addition, even with annihilation the stepsize is still limited by the smallest surviving dipole.

On the other hand, these kind of explicit methods have the advantage that there is no need to solve systems of non-linear equations which is usually necessary for implicit methods. This is a serious drawback for the backward Euler formula, but near equilibrium, theoretically, infinite stepsizes are possible which can compensate for the extra time but it is not as efficient when the system is far from it. Consequently, an \emph{in between} method would be preferred.

\section{The applied numerical scheme\label{sec:scheme_intro}}

%\textcolor{green}{
%\begin{itemize}
%\item trapéz súlyfaktorokkal, hogy miért kell a SF arra később visszatérünk. mindig stabil, ...
%\item az egyenlet megoldása, Newton Raphson ??? módszer, Jacobi mátrix, stb.
%\item hoppá ez így N\^3-ös -> szar. igen, de a kis időlépést a nagyon szűk dipólok okozzák -> %levágófüggvény a Jacobiban.
%\item súlyfaktorok belövése
%\item adaptív stepsize
%\end{itemize}
%}

\subsection{Weighted implicit trapezoidal scheme (WITS)}
\label{sec:weighted_trapezoidal_scheme}

We start by introducing the so-called weighted implicit trapezoidal scheme. When solving the $\dot x(t) = f(x(t),t)$ ordinary differential equation this scheme with the notation of Eq.~(\ref{eq:derivative}) reads as
\begin{align}
F(x^{k+1},x^k, x^{k-1}, \dots, t^k) = \frac{(1-d) f(x^k, t^k) + (1+d) f(x^{k+1}, t^{k+1})}{2} \label{eq:trapezoid_rule}
\end{align}
In the equation $d$ is a weight factor that tunes the system between a backward Euler ($d=1$) and a forward Euler ($d=-1$) method.

The advantage of the introduction of the weight factor $d$ is that if it is chosen properly it can improve the efficiency of the integration. As it will be seen later, the symmetric trapezoidal rule ($d=0$) is not optimal in situations where dislocations can \emph{jump over} their equilibrium positions, which may lead to spurious (though non-diverging) oscillations. This nonphysical behaviour can be removed by appropriate choice of $d$ (see section \ref{sec:weight_selection}) which ensures energy dissipation. It is noted that a similar approach was applied in the Parametric Dislocation Dynamics simulator \citep{ghoniem_parametric_2000}.

When applying the method to a dislocation system Eq.~(\ref{eq:derivative}) with Eq.~(\ref{eq:trapezoid_rule}) has to be solved for every dislocation. From now on the lower index of a symbol will identify the particular dislocation and upper index will mark the given time step. After introducing the function
\begin{equation}
    g_i(x_1,\dots,x_N) := x_i-\frac{(1+d_i)h}{2} f_i(x_1,\dots,x_N,y_1,\dots,y_N) - x_i^k-\frac{(1-d_i)h}{2}f_i^k,
\end{equation}
reorganizing the formula and inserting Eq.~(\ref{eq:x_eq_motion}) one obtains
\begin{align}
    g_i^{k+1} =& g_i(x_1^{k+1},\dots,x_N^{k+1}) =  x_i^{k+1}-\frac{(1+d_i)h}{2} f_i^{k+1} - x_i^k-\frac{(1-d_i)h}{2}f_i^k \notag \\
    =&x_i^{k+1}-\frac{(1+d_i)h}{2}s_i\sum_{j\neq i}^N s_j\tau_\text{ind}\left(x_i^{k+1}-x^{k+1}_j, y_i^{k+1}-y^{k+1}_j\right) \label{eq:g_intro}\\
    &-x_i^k-\frac{(1-d_i)h}{2}f_i^k=0, \notag
\label{eq:non_lin}
\end{align}
which is a system of non-linear equations that needs to be solved to get the coordinates $x_i^{k+1}$ of the dislocations at the new timestep (the $x_i^k$, and, thus, $f_i^k$, values are known from the previous timestep).

\subsection{Solving the non-linear equation system}

To solve Eq.~(\ref{eq:non_lin}) Newton-Raphson iteration method was chosen where we assume that the solution is close to the actual position. Hence a few iterations (in our case 2) are expected to deliver a close enough solution. If this was not the case the adaptive stepsize protocol will correct the error by switching to smaller stepsize. The initial guess for the solution is the coordinate vector from the previous time step 
%(\textcolor{orange}{the upper indices in the parenthesis indicate how many iterations were performed till the given value at the given time}):
\begin{align}
    x_i^{k+1,(0)} = x_i^k,
\end{align}
and an iteration step takes the form
\begin{align}
    \sum_{j=1}^N \left( x_j^{k+1,(n)} - x_j^{k+1,(n+1)} \right)J_{ij}^k = g_i^{k+1,(n)}, \label{eq:iteration_step}
\end{align}
where $x_i^{k+1,(n)}$ approximates $x_i^{k+1}$ as $n$ increases and $J_{ij}^k$ is the Jacobian matrix and is evaluated at timestep $k$. The latter reads as
\begin{align}
%J_{ij}=& \label{eq:def_jacobian}\\
J_{ij}^k=&\frac{\partial g_i}{\partial x_j}(x_1^k,\dots,x_N^k)=\begin{cases}1-\frac{1+d_i}{2}hs_i\sum_{j\neq i}^Ns_j\partial_x\tau_\text{ind}(x_i^k-x_j^k,y_i^k-y_j^k), & \text{if } i=j,\\
\frac{1+d_i}{2}hs_is_j\partial_x\tau_\text{ind}(x_i^k-x_j^k,y_i^k-y_j^k), & \text{else.}
\end{cases}\label{eq:jacobian_full}
\end{align}

\subsection{Selection of the weight factors\label{sec:weight_selection}}

As it was pointed out previously, it is very important to select the right $d_i$ values to avoid oscillations of dislocations around equilibrium positions because that would violate conservation of energy in the system. Firstly, let us consider the case of a dipole of Sec.~\ref{sec:dipole}. Close to equilibrium the equation of motion is approximately
\begin{align}
\dot x \approx -\frac1{\tau} x, \label{eq:exponential}
\end{align}
%of these that near the equilibrium point the stress field can be approximated by an exponential function, hence its derivative is Eq.~(\ref{eq:exponential}).
where $\tau = d^2$.
By applying the weighted implicit trapezoid rule on that equation and requiring $x(t)>0$ for every $t$ if $x(0)>0$, the ideal weight factor $d$ can be obtained and reads as
\begin{align}
d= \frac{1}{1+\frac{2}{h/\tau}}. \label{eq:weight_simple}
\end{align}
According to this formula if the timestep $h \ll \tau$ then $d\approx 0$, so one may use the second order symmetric trapezoidal scheme. If, however, $h \gg \tau$ then $d \approx 1$ is obtained meaning that the lower order backward Euler method must be used to avoid oscillations. This modification to the scheme is not only advantageous numerically, but also motivated by physical arguments. In the equation of motion Eq.~(\ref{eq:x_eq_motion}) overdamped dynamics are assumed, so, the total energy of the system may only decrease due to the strong dissipation. Any numerical oscillation around an equilibrium position would, in fact, introduce energy to the system, which could lead to unphysical phenomena. With the weight factors introduced this possibility can be avoided.

In order to generalize this approach to the system of $N$ dislocations it is useful to define matrix $\hat{\bm{A}^k}$ as
\begin{align}
A_{ij}^k=\begin{cases}
-hs_i\sum_{j\neq i}^Ns_j\partial_x\tau_\text{ind}(x_i^k-x_j^k, y_i^k-y_j^k), & \text{if } i=j, \\
hs_is_j\partial_x\tau_\text{ind}(x_i^k-x_j^k, y_i^k-y_j^k), & \text{else.}
\end{cases}\label{eq:A_matrix}
\end{align}
The $d_i^k$ weights for each dislocation can be then determined as:
\begin{align}
d_i^k=\begin{cases}
\frac{1}{1+\frac{1}{A_{ii}^k}}, & \text{if } A_{ii}^k > 0, \\
0, &\text{else,}
\end{cases}\label{eq:d_i_determined}
\end{align}
For a short dipole this exactly reproduces Eq.~(\ref{eq:weight_simple}). Finally, at a given timestep the Jacobian matrix $\hat{\bm{J}}^k$ (according to Eq.~(\ref{eq:jacobian_full})) can be computed as
\begin{align}
J_{ij}^k=\delta_{ij}+\frac{1+d_i^k}{2}A_{ij}^k
\label{eq:J_from_A}
\end{align}
which means that complexity of the method is not increased with the introduction of the weight factors $d_i^k$.

\subsection{Reducing the complexity of the implicit method\label{sec:solution}}

The time complexity of a single timestep for the explicit integration schemes in DDD simulations is $\mathcal{O}(N^2)$, since all the pair interactions have to be taken into account at every timestep. Implicit methods, as explained above in detail, are expected to perform better due to larger possible stepsizes that can be made with the same numerical precision. Yet, there is a fundamental issue with such schemes that prevent them from being superior to explicit schemes for large system sizes. Namely, the linear system of equation of Eq.~(\ref{eq:iteration_step}) has to be solved several times at every timestep. The complexity of such solution is $\mathcal{O}(N^3)$ because the Jacobian matrix is dense (its components, representing a dislocation pair, decay as $r^{-2}$ with the mutual distance $r$ between the dislocations) which makes this complexity also apply for the whole scheme. This means that there is always a critical system size where the larger obtainable stepsize compared to the explicit one cannot compensate for the longer runtime of a single timestep.

In this section this problem is addressed by the introduction of a special cutoff function during the calculation of the Jacobian matrix. The basic idea is that the stepsize constraint in explicit systems are posed by the shortest dipoles in the system. So, by considering only these short dipoles in an implicit manner would already increase significantly the possible stepsize, similarly to the annihilation procedure applied for explicit schemes. We, therefore, define the cut-off function as
\begin{align}
f(x,y)=\begin{cases}
1,&\text{if } x^2+y^2 < r_c^2,\\
\exp\left(-\frac{\left(\sqrt{x^2+y^2}-r_c\right)^2}{r_c^2}\right), & \text{else},
\end{cases}\label{eq:define_cutoff}
\end{align}
where $r_c$ is a cut-off parameter setting a lengthscale. With this the Jacobian reads as
\begin{align}
%J_{ij}=& \label{eq:def_jacobian}\\
J_{ij}(x_i,\dots,x_N)=&\begin{cases}1-\frac{1+d_i}{2}hs_i\sum_{j\neq i}^Ns_j f(x_i-x_j,y_i-y_j) \partial_x\tau_\text{ind}(x_i-x_j,y_i-y_j), & \text{if } i=j,\\
\frac{1+d_i}{2}hs_is_j f(x_i-x_j,y_i-y_j) \partial_x\tau_\text{ind}(x_i-x_j,y_i-y_j), & \text{else.}
\end{cases}\label{eq:jacobian_full_2}
\end{align}
The exponentially decaying function $f$ makes the values of the Jacobian at a certain distance smaller than the floating-point precision. In the present case if the result of the function is less than $10^{-16}$ the corresponding element is replaced by zero. Consequently, the Jacobian becomes a sparse matrix. It can be easily seen that number of non-zero elements is approximately $Na$, where $a$ does only depend on $r_c$ as long as $r_c \ll L = \sqrt{N}$ holds. To solve a system of equations which is described by a sparse matrix is significantly faster than that of a dense matrix with complexity under $\mathcal{O}(N^2)$ \citep{liu_h-lu_2010}, so the complexity of the whole method remains $\mathcal{O}(N^2)$ (since the computation of pair interactions is still necessary at every timestep) just like in case of the explicit methods.

A similar method was used in \citep{Sills_2014}, where elastic interactions between dislocation segments were only considered if they were closer than a specified length parameter (a hard limit), but the lack of continuity in the derivative can cause performance drop because the Newton-Raphson iteration method may not converge if the root is close to or at the discontinuity.

It is clear, that by introducing the cutoff function an artificial analytical error appears in the calculation. Since only the Jacobian is influenced by this modification the guesses of the Newton-Raphson method are affected during the solution and the results of the non-linear equations become less precise. This will not result in systematic error since the adaptive stepsize control (see below) monitors the introduced error and recomputes a step with decreased timestep if necessary.

To elaborate further on this point it is noted that if the cut-off parameter $r_c$ is 0 the method becomes exactly an explicit method (in particular, with two iteration cycles inside the Newton-Raphson method it becomes the explicit predictor-corrector method), while if $r_c$ approaches infinity it becomes a fully implicit method. So this $r_c$ can be used to \emph{tune} the scheme between explicit and implicit. Typically those dipoles (or dislocation pairs in general) are considered implicitly where dislocations are closer than $r_c$.

To summarize the above statements, two control parameters are present in the scheme: the weights factors which are fixed by the actual configuration of a dislocation system at a given time and the cut-off distance $r_c$ which can be selected at any given timepoint differently and is not restricted physically.

\subsection{Adaptive stepsize control}
A simple algorithm was used to control the precision of the solution described by the following procedure: With a given $h$ stepsize a step was calculated, after that the same timepoint from the original timepoint was reached by two $\frac{h}{2}$ sized steps. The difference between the coordinates of the dislocations at $t+h$ was calculated between the two, the largest absolute value of these differences is called the error $\varepsilon$. This is required to be under a given limit $\varepsilon_{\text{max}}$ (tolerance). If this condition is not fulfilled the results are marked as failed and need to be recalculated. Otherwise, the result from the two smaller steps is stored at the new $t+h$ timepoint. The new stepsize is determined by the following equation in both cases:
\begin{align}
    h_{\text{new}}=\begin{cases}
    0.9h\cdot\min(2, \frac{\varepsilon_{\text{max}}}{\varepsilon}), \text{if }\varepsilon \neq 0\\
    2h, \text{ else.}
    \end{cases}
\end{align}

\section{Implementation\label{sec:implementation}}
The method described above was implemented in c++ in a strictly non-parallel fashion, but the scheme can be parallelised in an efficient manner and ported to GPU as well. To solve the arising, previously described linear equation systems, the \mbox{UMFPACK} library \citep{Davis_umfpack} was used which is used in MATLAB as well to solve systems of linear equations described by a sparse matrix.

During the implementation special care was taken to store the Jacobian directly in sparse format to avoid unnecessary storage operations. This is very efficient if the matrix is indeed sparse, but if it is not the case (e.g.,~a cutoff is not applied or it is comparable to the system size) it comes with an overhead during element lookup. Hence, with the fully implicit method ($r_c = \infty$) faster results are possible with a different storage scheme.

The shear stress field of a dislocation was precalculated with periodic boundary conditions with good resolution on a grid \citep{bako_dislocation_2006}. Near the origin the exact value of Eq.~(\ref{eq:field}) was calculated but for greater distances linear interpolation between the precalculated values was performed. The $\partial_x \tau_\text{ind}$ field needed for the Jacobian was evaluated in a similar manner. This approach outperforms the image based field calculation in case of single thread computing on our available hardware.

A version of the implementation is available on github under GNU General Public License v3.0: \hyperlink{https://github.com/pgabor/sdddst}{https://github.com/pgabor/sdddst}

\section{Numerical results\label{sec:numerical_results}}
\subsection{Effect of the cut-off parameter on the denseness of the Jacobian}
As it was described in section~\ref{sec:solution} the cut-off parameter reduces the complexity of the problem by calculating the elements of the Jacobian only where the distance between two dislocations is under a certain threshold, thus, the Jacobian matrix becomes a sparse one. In order to quantify its sparseness, we generated random dislocation configurations and the Jacobians were calculated for each one with different cut-off parameters. The actual number of the different configurations are summarized in Tab.~\ref{tab:random_config_numbers}.

\begin{table}[ht]
    \centering
    \begin{tabular}{c|c}
        Number of dislocations $N$ & Number of configurations \\
        \hline
        64 & 10000 \\
        256 & 5000 \\
        1024 & 2000 \\
        4096 & 1000 \\
        16384 & 100
    \end{tabular}
    \caption{Number of different dislocation configurations for each dislocation system size for the density measurement of the Jacobian.}
    \label{tab:random_config_numbers}
\end{table}

\begin{figure}[ht!]
    \centering
    \includegraphics{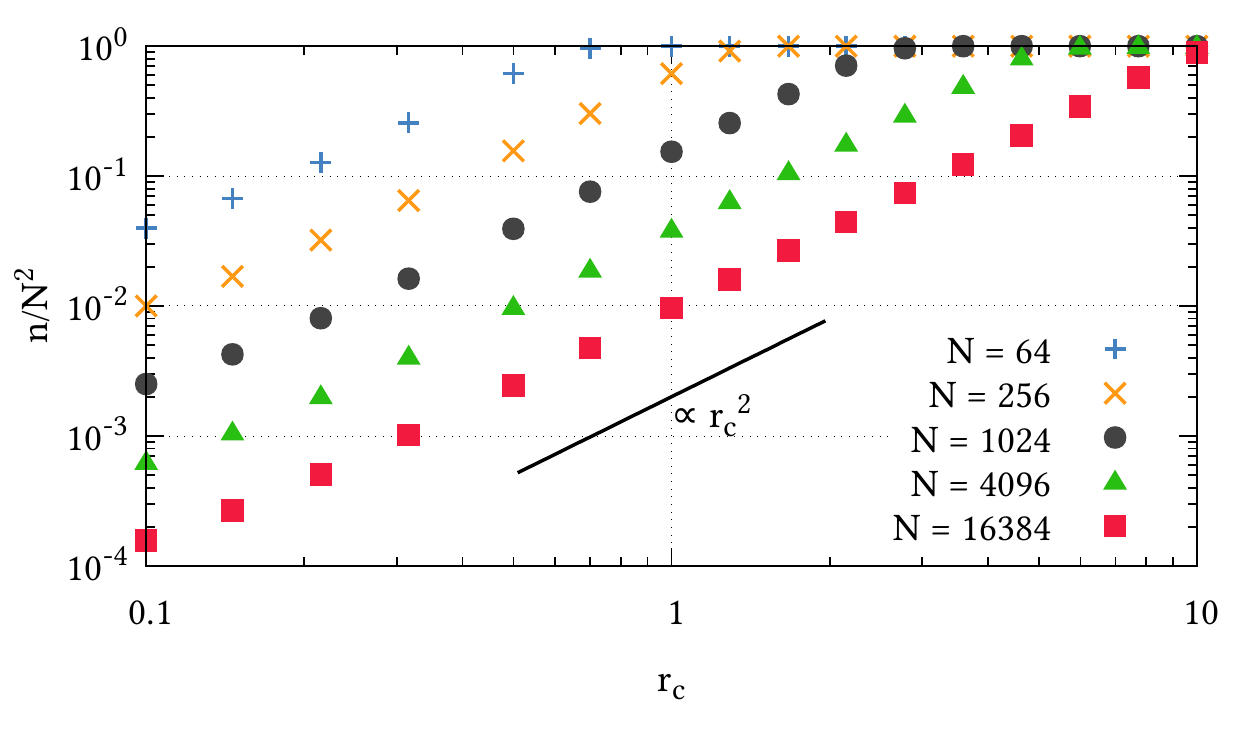}%}nonzeroratio_vs_cutoff.pdf}
    \caption{The ratio of the non-zero elements in the Jacobian matrix for different system sizes as a function off the cut-off distance $r_c$. $n$ stands for the number of the non-zero elements and $N$ is the number of dislocations in the system.}
    \label{fig:jacobian_density}
\end{figure}

The results can be seen in Fig.~\ref{fig:jacobian_density} where $n$ denotes the number of non-zero elements in the $N\times N$ sized Jacobian. As explained above in section~\ref{sec:solution} the number of non-zero elements is approximately $n = Na$, where $a$ is the average number of non-zero elements in a row. The latter is obviously proportional with the number of dislocations where the cut-off function $f$ returns a non-zero value, so, $a\propto r_c^2$. This leads to the result $n \propto N r_c^2$, which can be clearly observed in Fig.~\ref{fig:jacobian_density}. There are two limits to this behaviour: (i) For small $r_c$ values $n$ tends to $N$, since the diagonal elements do not vanish even for $r_c=0$ (and one arrives at an explicit method as explained previously) and (ii) for large $r_c$ values all matrix elements become non-zero, so $n$ saturates at $N^2$.

%The ratio of the non-zero elements increase proportionally with $r_c^2$ in the greatest part, but it can not reach zero, because the diagonal elements are going to be present even when the other elements are missing. As a result the lowest achievable density is $1/N$ what corresponds to an explicit method. An other observable phenomena that after a certain $r_c$ value the non-zero element ratio saturates at 1. This means that for the given system size the chosen $r_c$ value is large enough to handle each dislocation interaction in an implicit manner. Of course when all elements are present it can not lower the complexity of $\mathcal{O}(N^3)$ which is associated with the solution of the linear equation system.

These results clearly show that with the right choice of $r_c$ a sparse matrix can be obtained that with increasing $N$ becomes even sparser. Hence, the complexity may go below $\mathcal{O}(N^2)$ to solve the linear equation system. In the next chapter we compare the runtime and the achieved stepsize for different cut-off parameters for dislocation systems of various sizes.

\subsection{Effect of the cut-off parameter on the runtime of the simulations}

\begin{figure}[ht!]
    \centering
    \begin{picture}(0,0)
    \put(26,218){\sffamily{(a)  $N=256$}}
%    \put(90,180){$N=256$}
    \end{picture}
    \includegraphics{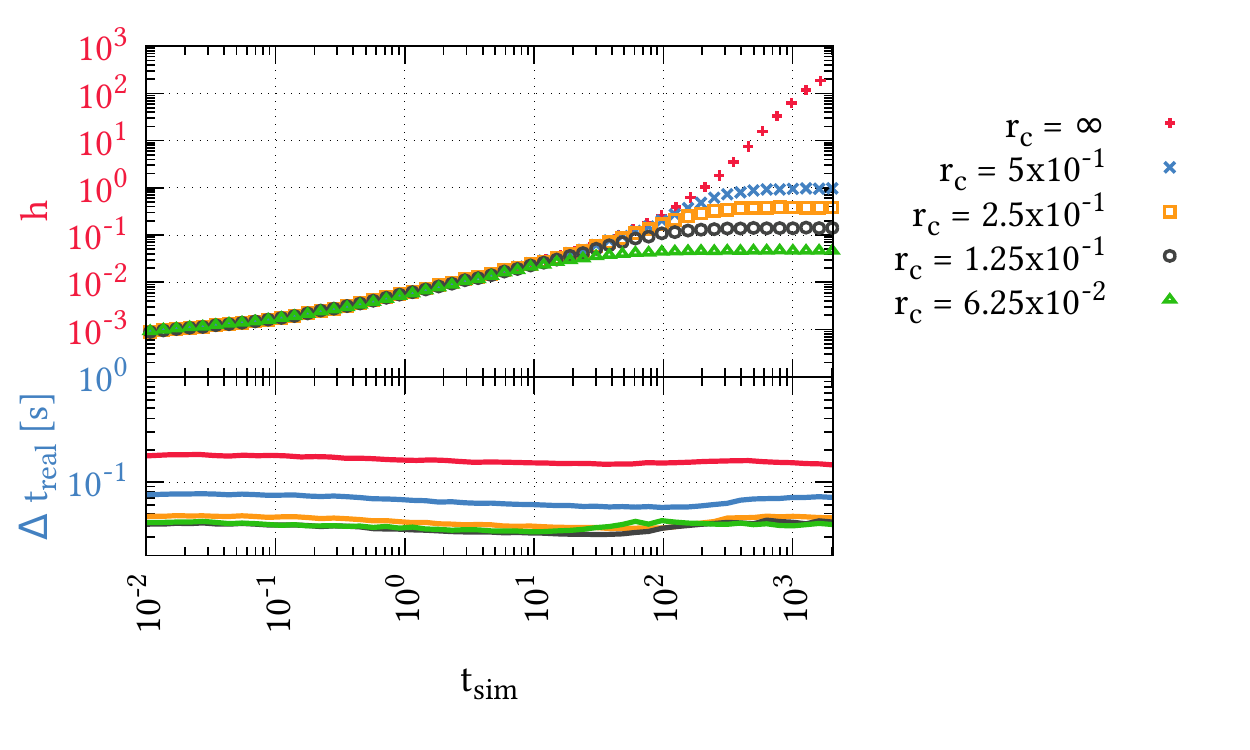}%256_delta_t_plot.pdf}
    
    \begin{picture}(0,0)
    \put(26,218){\sffamily{(b) $N=1024$}}
%    \end{picture}
%    \begin{picture}(0,0)
%    \put(87.5,176.5){$N=1024$}
    \end{picture}
    \includegraphics{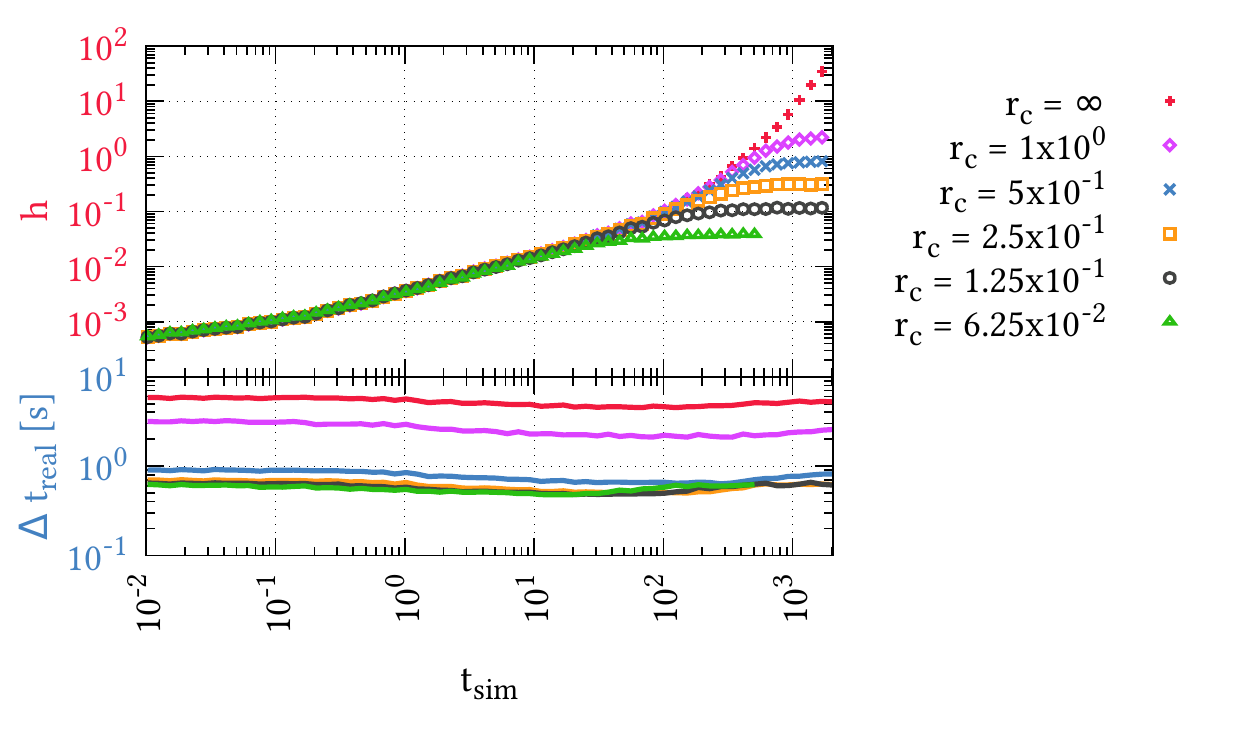}%1024_delta_t_plot.pdf}
    \caption{Dependence of the stepsize ($h$) and the spent wallclock time ($\Delta t_{\text{real}}$)  between to successful timesteps as a function of the simulation time ($t_{\text{sim}}$) for different $r_c$ values and (a) $N=256$ dislocations and (b) $N=1024$ dislocations. }
    \label{fig:delta_t}
\end{figure}
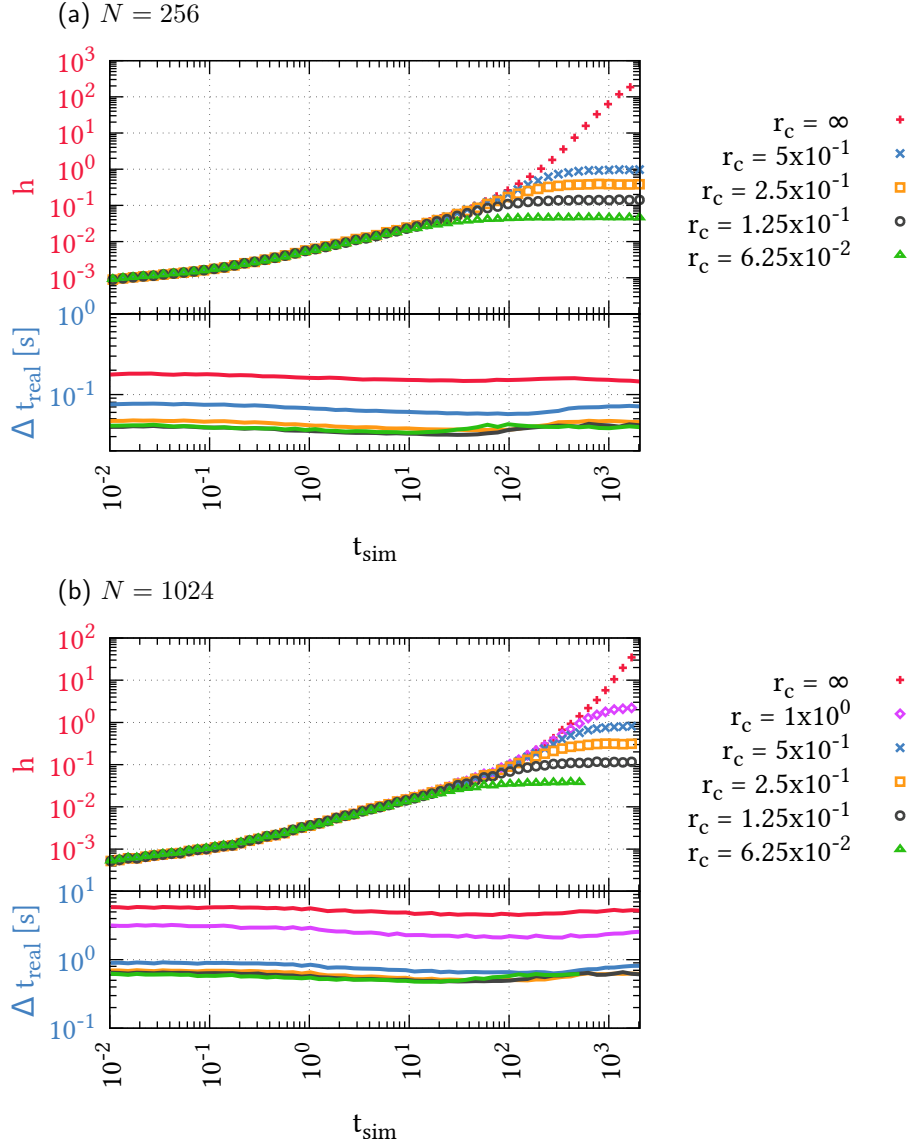

The relaxation of different initially random dislocation systems was performed with 1000 realizations for $N=256$ dislocations where the tolerance was $1.6\cdot10^{-5}$ and with 500 realizations for $N=1024$ dislocations with $3.2\cdot10^{-5}$ as maximal tolerance. Different $r_c$ parameters were used but $r_c$ was kept fixed during every simulation run. In each case the spent wallclock time ($\Delta t_{\text{real}}$) and the achieved stepsize ($h$) was measured between every successful subsequent simulation step and the actual total simulation time ($t_\text{sim}$) was also recorded.
The resulting data was sampled on an equidistant interval on logarithmic scale and the corresponding data were averaged for the different realizations (same dislocation number $N$ and parameter $r_c$). If no exact datapoint has been found, linear interpolation was used between the two closest ones. The results are shown in Figs.~\ref{fig:delta_t}(a) and \ref{fig:delta_t}(b) for the $N=256$ and $N=1024$ case, respectively.

%We relaxed 1000 different random dislocation systems with 256 dislocations and 500 with 1024 dislocations with a fix $r_c$ parameter during the whole simulation. In each case the spent wallclock time ($\Delta t_{\text{real}}$) was measured between every successful simulation step and was recorded along the achieved stepsize ($h$) and actual time inside the simulation ($t_\text{sim}$). We sampled the resulting data on an equidistant interval on logarithmic scale and averaged the corresponding data (same dislocation number $N$ and $r_c$ parameter). Where no exact datapoint has been found, linear interpolation was used between the two closest ones. The results are Fig.~\ref{fig:256_delta_t} and Fig.~\ref{fig:1024_delta_t} for the $N=256$ and $N=1024$ case in order.

%\begin{figure}[ht]
%    \centering
%    \includegraphics{1024_delta_t_plot.pdf}
%    \caption{This figure shows the dependence of the stepsize ($h$) and the spent wallclock time ($\Delta t_{\text{real}}$) used for computation between to successful timesteps as a function of the time inside the simulation ($t_{\text{sim}}$) for different $r_c$ values. The number of the dislocations inside a simulation was 1024. The $x$-axis is common for both parts of the figure, but the $y$-axis is different. This is indicated by blue and red colors. The same color indicates the same $r_c$ parameter in case of the data.}
%    \label{fig:1024_delta_t}
%\end{figure}

According to the figures, in line with the expectations,  $\Delta t_\text{real}$ increases significantly with increasing cut-off $r_c$. In addition, it hardly changes during the course of the relaxation. This means that the sparseness of the Jacobian in nearly constant so it is not much affected by the formation of local correlated configurations seen in Figs.~\ref{fig:relaxation_figure}(b) and \ref{fig:relaxation_figure}(c). This also implies that the results of the previous section conducted for random dislocation ensembles remain approximately valid even after relaxation. On the other hand, the achieved average timestep $h$ increases significantly as the final equilibrium configuration is approached. More specifically, in the first part of the relaxation, where the system exhibits the largest activity, the achieved stepsize does not depend on $r_c$ for the values considered. However, as the activity ceases, the timesteps saturate at values that increase significantly with $r_c$ and even seemingly diverge for $r_c = \infty$.

%The figures show no direct connection between $\Delta t_\text{real}$ and $h$. While $\Delta t_\text{real}$ is almost the same for a given $r_c$ during the whole simulation, $h$ shows a different behaviour. This behaviour has some important consequences. Regardless of the $r_c$ choose, for the same $r_c$ the $\Delta t_\text{real}$ is going to be the same in average no matter of the activity of the dislocations, however the relationship is not unique as it can be seen for large $t_\text{sim}$.

%The other observation is that at the beginning when the originally random dislocation configuration shows the largest activity, regardless of $r_c$ the achieved stepsize is almost the same. When the system reaches a more quiet regime, the differences start to appear. Larger $r_c$ values (inside the non-saturated $n/N^2$ region, as it was discussed in the previous chapter) result in larger $h$ values, which is expected because a more dense Jacobian matrix contains more information about the underlying system.

\begin{figure}
    \centering
    \begin{picture}(0,0)
    \put(30,215){\sffamily{(a)} $N=256$}
    %\put(90,180){$N=256$}
    \end{picture}
    \includegraphics[width=0.5\textwidth, trim={2.5cm 0 2.5cm 0}, clip]{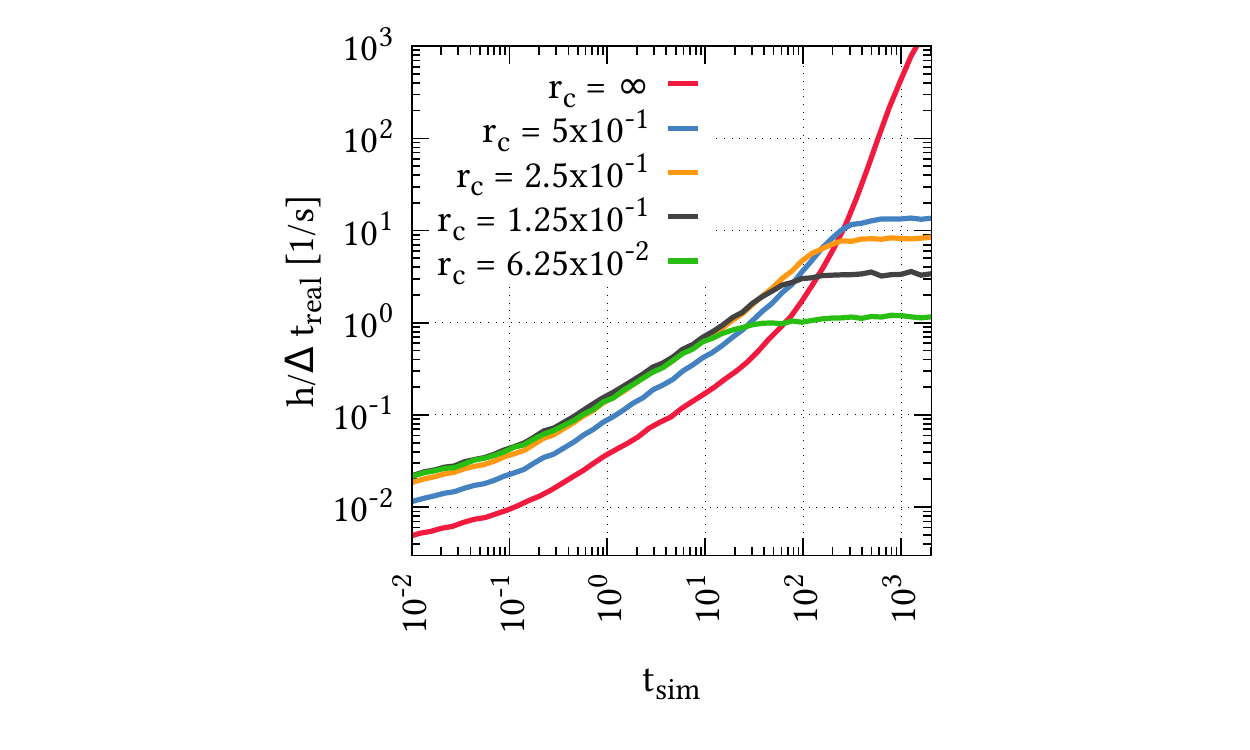}%256_simulation_step_eff_normal.pdf}
    \hspace{-0.5cm}
    \includegraphics[width=0.5\textwidth, trim={2.5cm 0 2.5cm 0}, clip]{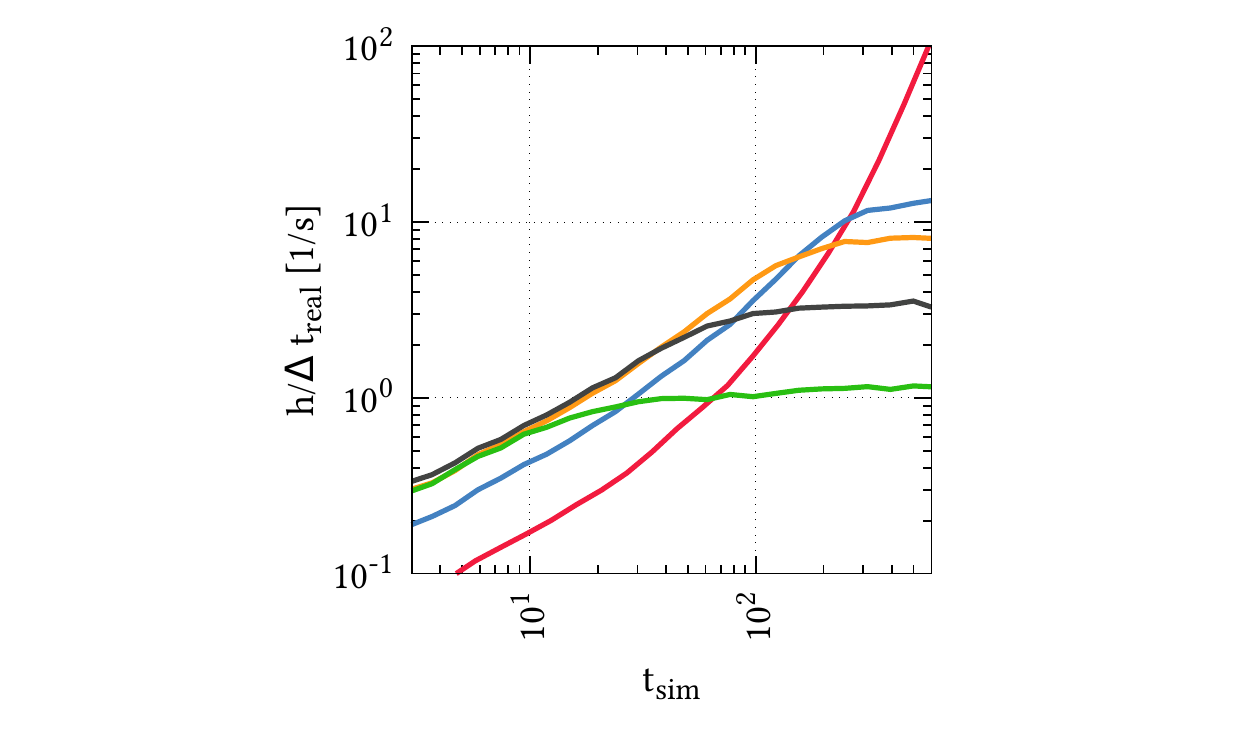}%256_simulation_step_eff_zoomed.pdf}
    
    \begin{picture}(0,0)
    \put(30,215){\sffamily{(b)} $N=1024$}
    %\put(90,180){$N=256$}
    \end{picture}
    \includegraphics[width=0.5\textwidth, trim={2.5cm 0 2.5cm 0}, clip]{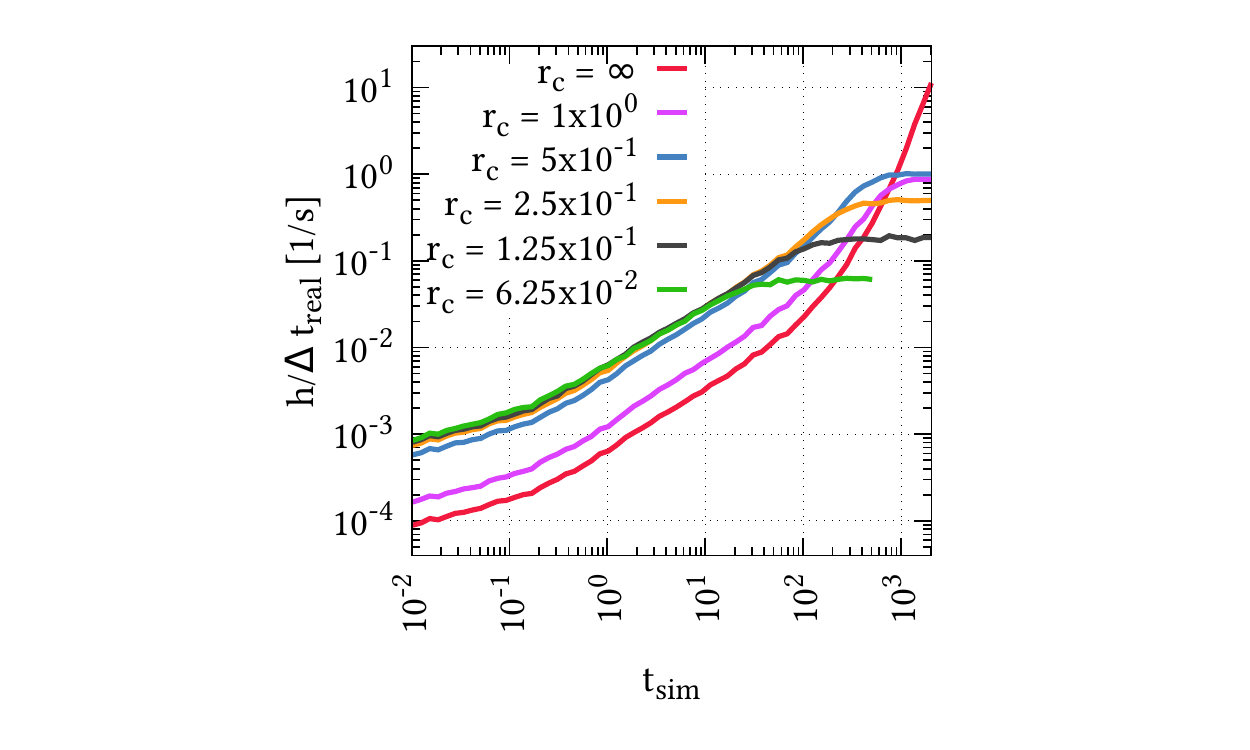}%1024_simulation_step_eff_normal.pdf}
    \hspace{-0.5cm}
    \includegraphics[width=0.5\textwidth, trim={2.5cm 0 2.5cm 0}, clip]{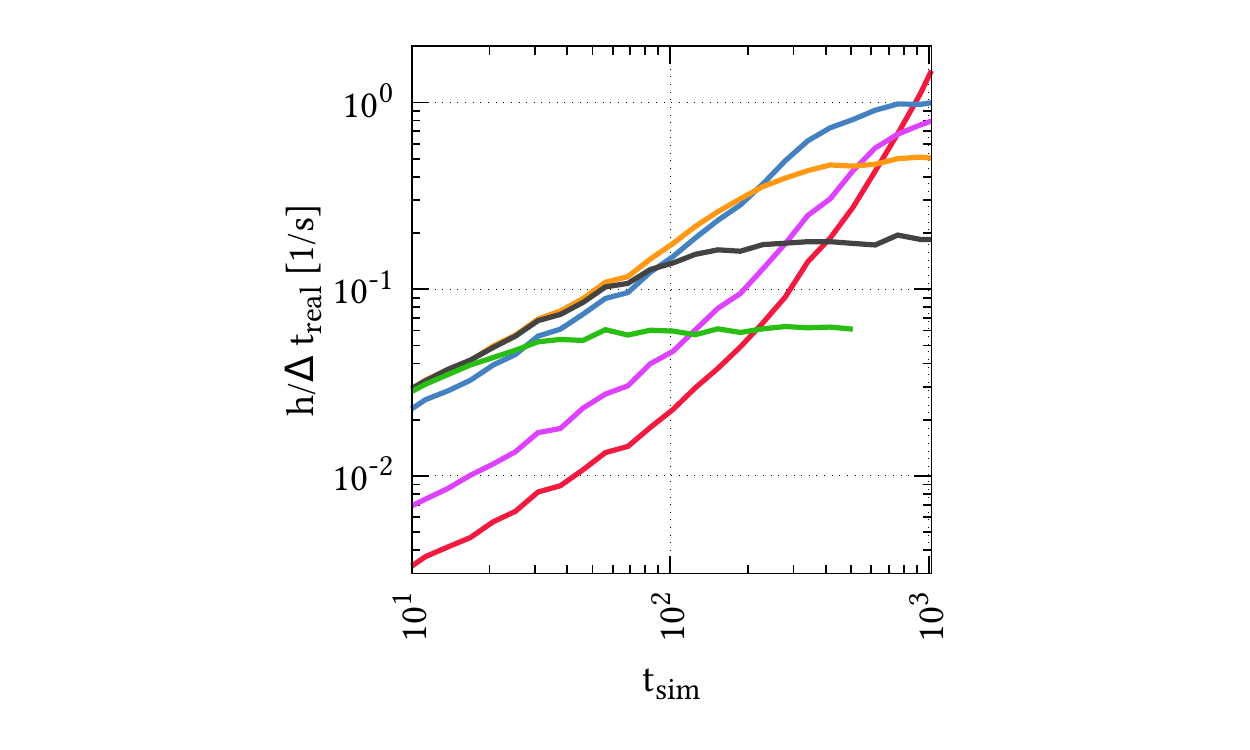}%1024_simulation_step_eff_zoomed.pdf}
    \caption{The efficiency $h/\Delta t_\text{real}$ of the new method for different $r_c$ values as a function of the simulation time $t_\text{sim}$ for (a) $N=256$ and (b) $N=1024$ dislocations. 
    The left panel plots the whole simulation time interval while the right one magnifies the part where the curves cross each other.}
    \label{fig:efficiency}
\end{figure}

Based on the previous observations the efficiency, defined as the simulation time advance under unit wallclock time, i.e., $h/\Delta t_\text{real}$, was also calculated for both system sizes. According to the plots of Fig.~\ref{fig:efficiency} at the beginning in the active regime, due to the smaller computational cost, the efficiency is better for small $r_c$ values. However, as the system gets closer to equilibrium a saturation in efficiency is seen and here the efficiency gets better for larger $r_c$ due to the increased timestep. The efficiency without a cut-off ($r_c=\infty$) seems to diverge in equilibrium.

%\begin{figure}
%    \centering
%    \subfloat[Unzoomed]{\includegraphics[width=0.5\textwidth, trim={2.5cm 0 2.5cm 0}, clip]{1024_simulation_step_eff_normal.pdf}}
%    \subfloat[Zoomed]{\includegraphics[width=0.5\textwidth, trim={2.5cm 0 2.5cm 0}, clip]{1024_simulation_step_eff_zoomed.pdf}}
%    \caption{The efficiency of the new method for different $r_c$ values as a function of the time inside the simulation ($t_\text{sim}$). The left the figure contains the whole simulation time interval while the right one magnifies the part where the curves interchange their position. On both figures the color is the same for the same $r_c$ value. These curves correspond to the simulations with $N=1024$ dislocations.}
%    \label{fig:1024_efficiency}
%\end{figure}

Every considered choice of $r_c$ leads to a significant decrease in total runtime compared to explicit methods. As seen in Fig.~\ref{fig:efficiency}, smaller $r_c$ values are more favourable in active (small $t_\text{sim}$) regions and larger ones gradually become more efficient as activity ceases ($t_\text{sim}$ increases). It is interesting to note, that for $N=1024$ dislocations the choice of $r_c=1$ is never the most efficient. The reason behind this behaviour turned out to be that at this sparsity level (approx.~20\% according to Fig.~\ref{fig:jacobian_density}) the used linear solver (UMFPack) started to use a different kind of strategy than for sparser cases. So, such details of the applied algorithms may also need to be taken into account for choosing the appropriate value for parameter $r_c$.

%From the previous observations the conclusion is that certain level of activity inside the system requires different $r_c$ values and as a thumb of rule the lower the activity is, higher $r_c$ is needed, but the constraints of the system size has to be considered as well, as it was presented in case of Fig.~\ref{fig:jacobian_density}. One can also notice for the larger $N=1024$ systems that the previous statement not one hundred percent true. The efficiency curve of $r_c=1$ never reaches a point where it is above the others and it needs more computation time as well relatively than the smaller $r_c$ values. This is the result of the fact that the sparsity reached a level where the used linear solver (UMFPack) started to use a different kind of algorithm. We verified this effect outside of the simulation context with solving linear equation systems with with different matrix sparseness. So this effect needs to be considered as well for choosing the right $r_c$ parameter.

\subsection{The importance of the weighting}

\begin{figure}
    \centering
    \begin{picture}(0,0)
    \put(30,215){\sffamily{(a)}}
    \end{picture}
    \includegraphics[width=0.5\textwidth, trim={2.5cm 0 2.5cm 0}, clip]{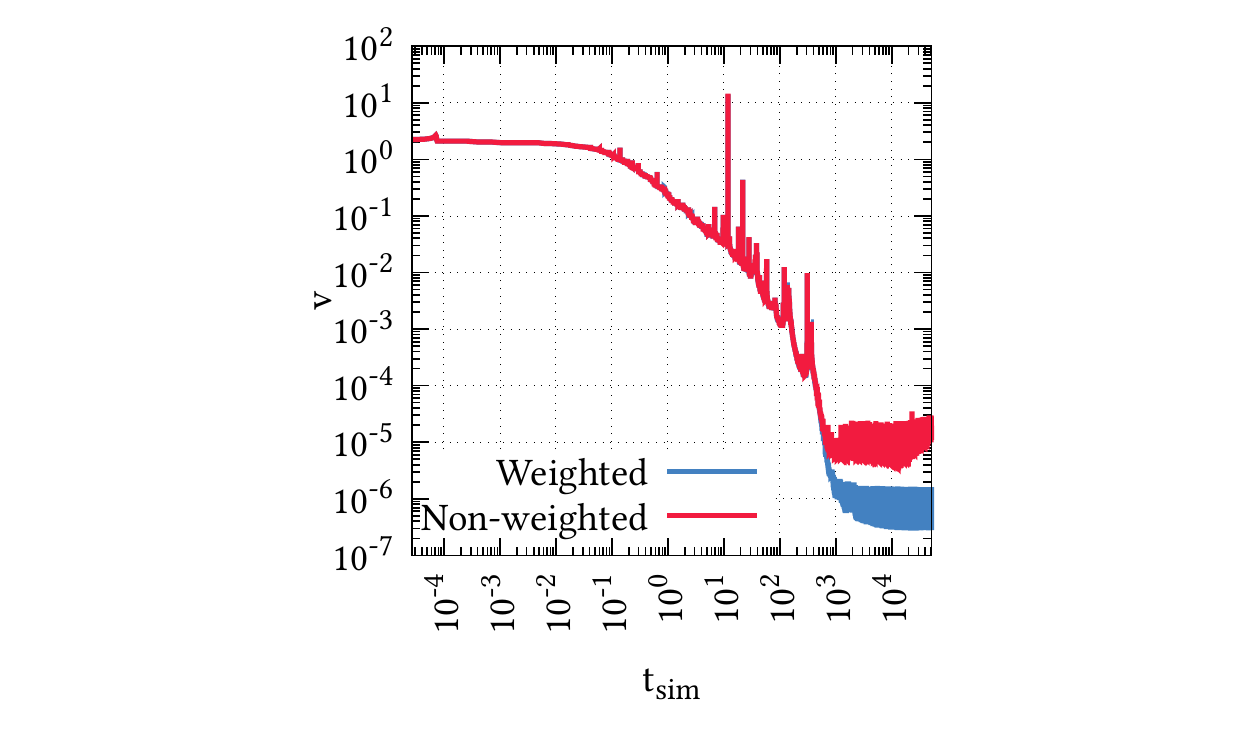}%cutoff_relax_speeds.pdf}
    \hspace{-0.5cm}
    \begin{picture}(0,0)
    \put(30,215){\sffamily{(b)}}
    \end{picture}
    \includegraphics[width=0.5\textwidth, trim={2.5cm 0 2.5cm 0}, clip]{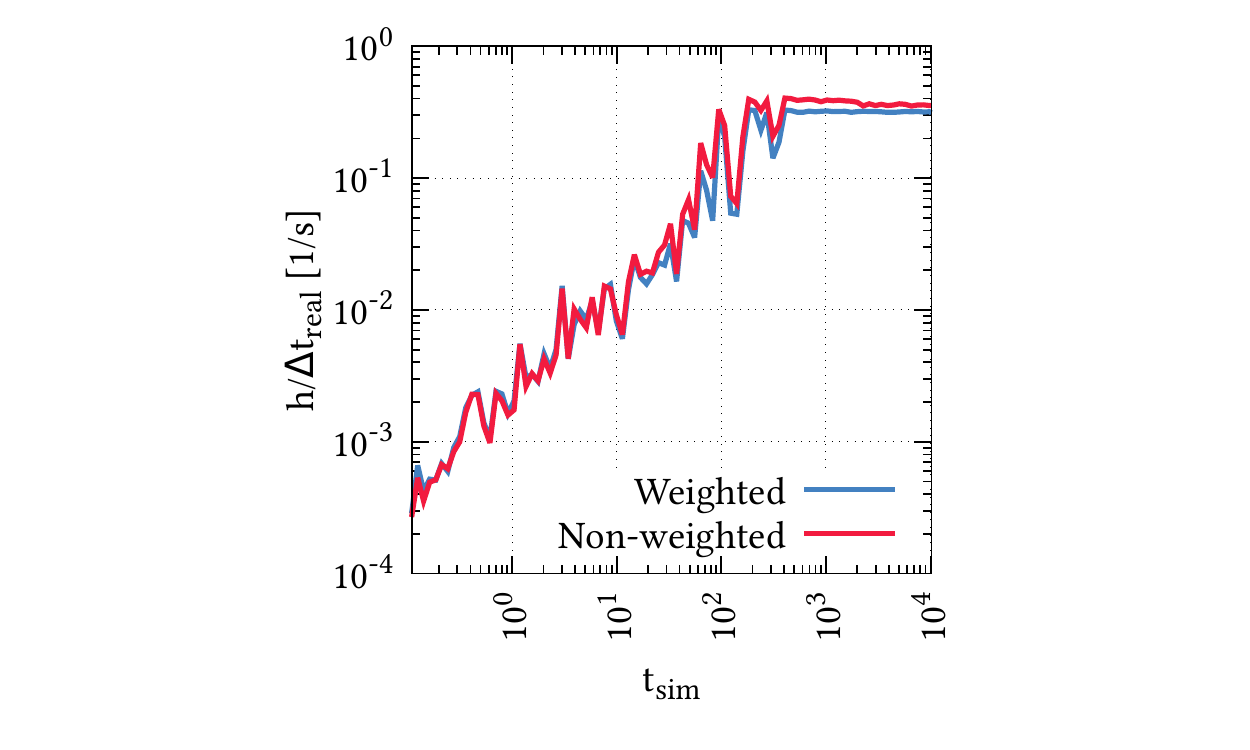}%cutoff_efficiency.pdf}
    \caption{(a) The average dislocation speed as a function of simulation time with the implicit trapezoidal scheme with and without weighting with $r_c=0.25$ and $N=1024$. (b) The corresponding efficiency. %\textcolor{red}{(BTW: hogyan nézne ki a bal oldali ábra $r_c = \infty$ esetén?)}
    }
    \label{fig:effect_of_wights}
\end{figure}

In order to decrease numerical oscillations around the equilibrium position of dislocations, an effect that violates conservation of energy in this system, the WITS was introduced in Sec.~\ref{sec:weighted_trapezoidal_scheme}. To demonstrate the advantage of this scheme relaxation simulations were performed with the same original dislocation configuration of $N=1024$ dislocations with the weighted and with the symmetric (i.e, $d_i=0$ for all dislocations) trapezoidal scheme. The cut-off was set to $r_c=0.25$ in both cases. According to the average velocity-time profiles seen in Fig.~\ref{fig:effect_of_wights}(a) there is no noticeable difference in the active regime, so, the motion of the dislocations is identical in the two cases. However, when motion stops and only numerical noise is seen in the final regime the level of noise is more than an order of magnitude lower in the case of the weighted scheme. This better performance does not require higher computational cost as seen in Fig.~\ref{fig:effect_of_wights}(b), where the efficiency of the two methods, defined above, are compared.

It is noted that further decrease of the numerical noise is possible with increasing the cut-off parameter $r_c$. As mentioned before, with the fully implicit method ($r_c = \infty$) the numerical noise can be decreased to the floating point precision but with the cost of significantly higher computational time of a single timestep. With a finite value of $r_c$ the simulation stepsize $h$ will converge to a finite value in equilibrium (see Fig.~\ref{fig:delta_t}), signaling the presence of some numerical noise. The fact that the WITS scheme is able to decrease significantly this noise with preserving the same $r_c$ value (and, at the same time, not increasing the computational cost of a single timestep) proves its advantageous properties compared to the traditional symmetric trapezoidal scheme.

%In the following through an example we present why the weighting what was introduced chapter~\ref{sec:weight_selection} is important. The example is based on simulations which have been done with the same original dislocation configuration with $N=1024$ dislocations, but with or without the weighting. The cut-off was $r_cm=0.25$ in both case. The relax curve of the simulations and the efficiency of the methods are shown on Fig.~\ref{fig:effect_of_wights}.

%The figure shows that without weighting the average speed starts to increase in an oscillating way while this rising effect is missing in case of the weighted approach. The oscillation is present because we introduced error in the calculation of Jacobian with the introduced cut-off function, hence the result can not describe the equilibrium better with the given precision constraints what is enforced by the adaptive stepsize control. This is the cause of the efficiency saturation as well. Interesting that the efficiency of the not weighted method is better but the difference is negligible considering the one magnitude better result.

\subsection{Comparison with an explicit method}

In this section the performance of the WITS method described above and an efficient 4.5th order Runge-Kutta (RK45) explicit scheme is compared. As it was mentioned above, annihilation is usually introduced in RK45 with a prescribed threshold distance to increase the efficiency. To be able to analytically compare the two methods annihilation was not performed with RK45, instead, 1000 random dislocation configurations with 256 dislocations were created and the one with the highest minimal possible dipole size was chosen as the starting configuration. During the relaxation the smallest dipole formed had a distance of $1.9 \times 10^{-3}$ in dimensionless coordinates, that is, approx.~$0.2 \%$ of the average dislocation spacing. Assuming, for instance, $10^{12}$ m$^{-2}$ as statistically stored dislocation density, this distance would be 2 nm, which is certainly above the annihilation distance at small temperatures. Therefore, in a such a scenario, introducing annihilation with realistic parameters would not speed up the simulation at all.

The average velocity vs.~time plots are plotted in Fig.~\ref{fig:explicit_implicit_compare} demonstrating that the motion of the dislocations is identical with the different methods. For the WITS method $r_c = 0.5$ was chosen. It is noted that to achieve this overlap a significantly smaller tolerance parameter had to be set for explicit method than for implicit ($1.6\cdot10^{-9}$ and $1.6\cdot10^{-5}$, respectively ). Further decrease of the tolerance parameter did not affect the relaxation curves of Fig.~\ref{fig:explicit_implicit_compare}.

The simulation runtime is remarkably different for the two methods, as shown in Fig.~\ref{fig:spent_time}. The total runtime needed to reach $t_\text{sim} = 614.4$ is plotted for the RK45 and the WITS method. Whereas to finish the simulations the RK45 needed more that 60 days it took only between 5 and 40 minutes for the WITS depending on the parameter $r_c$.

{\color{orange}
%As it was already mentioned, for DDD the 4th order explicit Runge-Kutta method is often used along with some dislocation dipole elimination distance. We compared the results of such a simulator with the method which was introduced by us. We created 1000 random dislocation configurations with 256 dislocations, and selected one where the minimal dipole size what can form during the simulation was the highest (hence decreasing the arising time step limitation for the Runge-Kutta method). We let the Runge-Kutta based simulators to run the simulations for approximately 70 days with different precision values, by a magnitude difference between them.

%We compared the average speed results for the here presented method and between the results from the Runge-Kutta methods on Fig.~\ref{fig:explicit_implicit_compare}. The cut-off was $r_c=0.5$ during the simulation in case of the new method and it took approximately 5 minutes to run which is several magnitudes of runtime difference. On Fig.~\ref{fig:spent_time} we presented, how much time was needed for different simulations with different $r_c$ values to reach the same $t_\text{sim}=614.4$ where the Runge-Kutta method was shut down.

\begin{figure}
    \centering
    \includegraphics{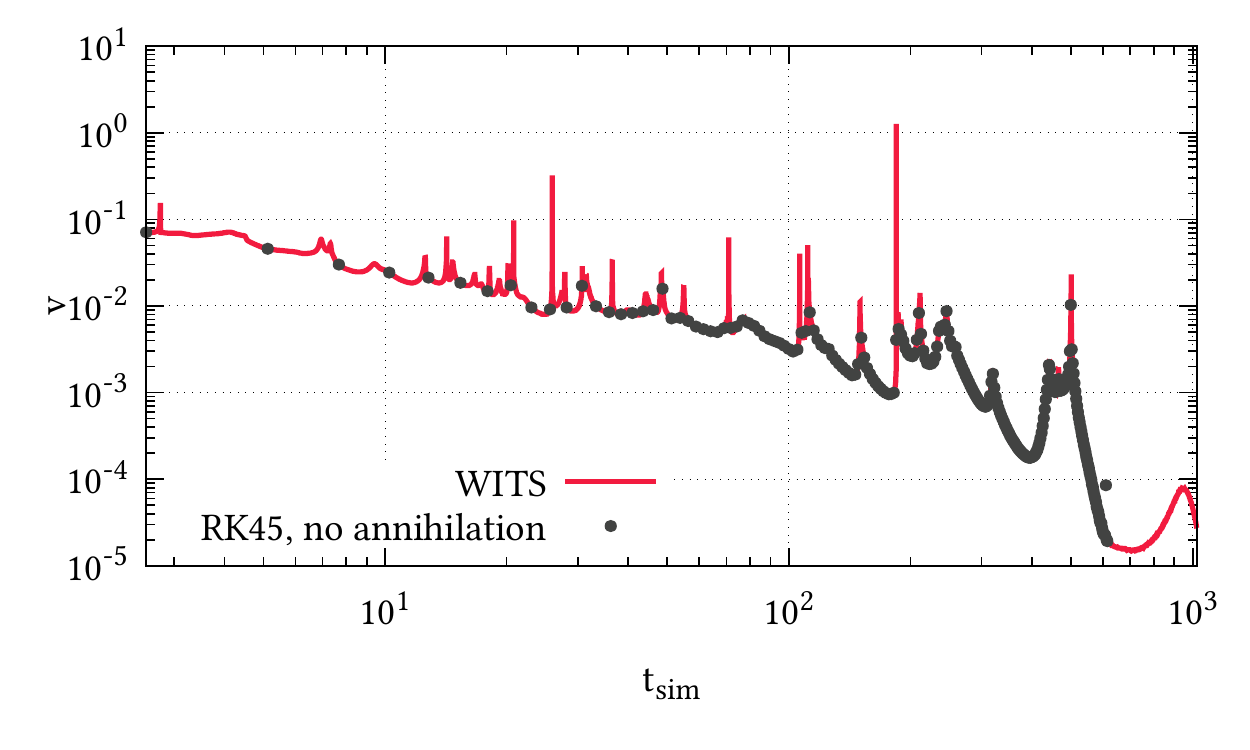}%256_explicit_implicit_speed.pdf}
    \caption{Comparison of average speed $v$ obtained by relaxing the same dislocation configuration containing $N=256$ dislocations with the 4.5th order explicit Runge-Kutta (RK45) method without dislocation annihilation and $1.6\cdot10^{-9}$ as tolerance and the implicit method introduced in this paper with $1.6\cdot10^{-5}$ tolerance and cut-off parameter $r_c=0.5$.}
    \label{fig:explicit_implicit_compare}
\end{figure}

\begin{figure}
    \centering
    \includegraphics{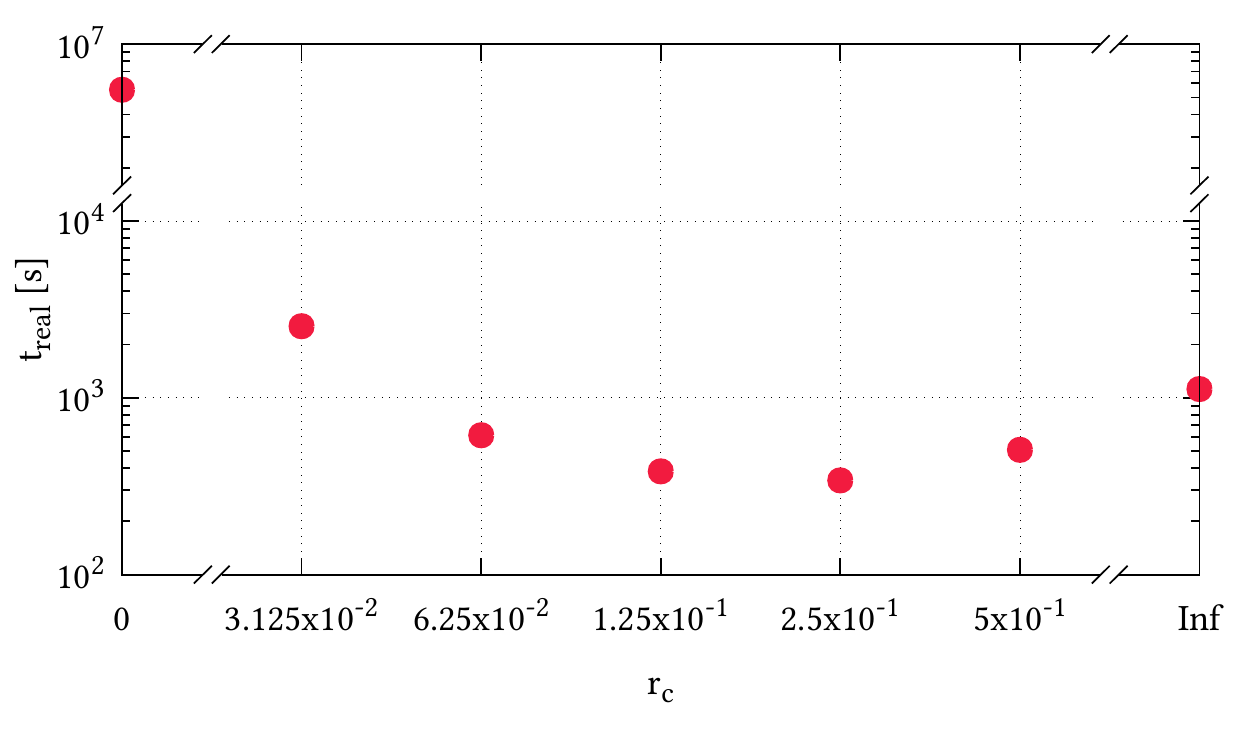}%runtime_00020.pdf}
    \caption{Wallclock time what was used to reach $t_\text{sim}=614.4$ in case of different methods. $r_c=0$ was obtained from the same Runge-Kutta based simulation what was used for Fig.~\ref{fig:explicit_implicit_compare}. The other datapoints correspond to the method described in this paper. In case of $r_c=\infty$ no cutoff function was used for the calculation of the Jacobian.}
    \label{fig:spent_time}
\end{figure}

%The maximum coordinate error for the Runge-Kutta method which was used to create Fig.~\ref{fig:explicit_implicit_compare} was set to $10^{-10}$ because from the simulations what were run this was the first what recreated the same average velocity development what we got in case of the new method with $10^{-6}$ precision. The simulations what used higher values were not able to reproduce it. This means the new approach is more precise in case of more loose precision control as well besides its better efficiency.

}
\section{Summary, conclusions and outlook\label{sec:summary}}

Discrete dislocation dynamics simulations play a central role in today's computational materials research as they can be used to model plastic deformation in crystalline materials without major approximations or assumptions in most cases. Unfortunately its application is strongly limited in achievable sample size (typically few microns) and duration (typically few microseconds) because of the long-range nature of stress fields individual dislocations create. These stress fields do not only decay slowly at far distances but also diverge at the core of the dislocation. As described in this paper these features make the equations of motion of the dislocations a \emph{stiff} set of differential equations. So far mostly explicit schemes were employed in DDD simulations where strong upper limit applies to the achievable timestep even in equilibrium. In particular, the maximum possible timestep is basically determined by the smallest dislocation dipole in the system even if the dipole is in its equilibrium, stationary state. To weaken this constraint dislocation annihilation is usually introduced, where the smallest dipoles are removed from the system thus increasing the stepsize.

Since these type of stiff equations can generally be more efficiently solved with implicit schemes, in this paper a novel implicit method was introduced with adaptive stepsize control to solve the differential equations governing the motion of dislocations. A weighted trapezoidal scheme was employed with specific weight factors in order to decrease oscillations around equilibrium and resulting numerical noise. A cut-off parameter $r_c$ was also introduced in the Jacobain matrix with the following properties:
\begin{itemize}
    \item At $r_c = \infty$ (that is, without using the cut-off functions) the numerical noise can be decreased down to floating point precision and the simulation timestep at equilibrium can practically diverge. These optimal properties are somewhat flawed by an increased computational cost of a single timestep because the Jacobian matrix is dense due to the long-range nature of interactions. In fact, the complexity of computing a single timestep is larger than that of the explicit methods (being $\mathcal O(N^2)$), so for large systems the implicit method only close to equilibrium may be more efficient than traditional explicit methods.
    \item At finite cut-off $r_c$ an in between method is obtained: On the one hand, it makes the Jacobian sparse thus decreasing the complexity of a timestep to $\mathcal O(N^2)$ and, on the other hand, it still allows the stepsize to be significantly larger than that of explicit schemes. It, in fact, acts similarly to the annihilation distance of explicit methods: dipoles with distance smaller than $r_c$ are treated implicitly (and thus not limiting the timestep) whereas those with distance larger than $r_c$ are computed explicitly and do limit the timestep even at equilibrium. The advantage is, that the small dipoles here do not have to be annihilated, and their dynamics are still solved with high precision. It was found that the runtime of the new method was approximately 4 orders of magnitude lower than that of the explicit method with a realistic annihilation distance.
\end{itemize}

The results obtained indicate that when activity was high in the system a smaller $r_c$ value was the most efficient and as activity ceased an increasing value of $r_c$ showed better performance. Future work will aim at developing an algorithm that dynamically changes the value of $r_c$ based on the dynamic properties of the system. Such a method could further improve the efficiency of this scheme. In addition, significant increase in computational speed can be expected from porting the source code to GPU, which is also relegated to future work.

In this paper the performance of the WITS was demonstrated on relaxation simulations where dislocations start from initially random configurations. After an initial high activity period dislocations gradually slow down as they approach equilibrium. The motivation behind studying such simulations was that dislocation dynamics is usually an intermittent process characterized by short high activity periods (strain bursts or dislocation avalanches) and long quiescent periods in between \citep{Miguel2001, dimiduk_scale-free_2006}. The relaxation process contains both limits: a very active initial phase together with an absolutely stationary state at large simulation times. An optimal numerical method should be efficient in both regimes, and according to the results presented so far, the implicit scheme described here fulfills this criterion.

The proposed scheme, thus, successfully addresses one of the main numerical challenges of DDD simulations, namely, that due to diverging and long-range stress fields the governing differential equations are stiff. The advantage of the newly developed scheme is that it is relatively simple and flexible and that it is not constrained to the specific 2D model used in this paper but can be easily generalized for more complex 2D or 3D DDD simulations given the Jacobian matrix can be analytically calculated.

\section{Acknowledgements}
Fruitful discussions with Istv\'an Groma are gratefully acknowledged.
This work was completed in the ELTE Institutional Excellence Program (1783-3/2018/FEKUTSRAT) supported by the Hungarian Ministry of Human Capacities. The present work was supported by the National Research, Development and Innovation Fund of Hungary (contract numbers: NVKP\_16-1-2016-0014, NKFIH-K-119561, NKFIH-KH-125380).
GP is also supported by the ÚNKP-19-3-I New National Excellence Program of the Hungarian Ministry for Innovation and Technology. PDI is also supported by the ÚNKP-18-4 New National Excellence Program of the Hungarian Ministry of Human Capacities and by the János Bolyai Scholarship of the Hungarian Academy of Sciences.

\bibliography{bibliography}

\end{document}